%% file: main.tex
\documentclass[journal,compsoc]{IEEEtran}
\IEEEoverridecommandlockouts

\usepackage[english]{babel}
\usepackage[utf8x]{inputenc}
\usepackage[T1]{fontenc}

\usepackage{amsmath}
\usepackage{graphicx}
\usepackage[colorinlistoftodos]{todonotes}
\usepackage[colorlinks=true, allcolors=blue]{hyperref}
\usepackage{enumitem}
\usepackage{pgfplots}
\usepackage{tikz,amsmath,amssymb,bm,color}
\usetikzlibrary{circuits.logic.US,circuits.logic.IEC,fit}

\newcommand{\codefn}[1]{{\footnotesize\textsf{#1}}}
\newcommand{\hide}[1]{}
\newcommand{\anon}[2]{#2}

\title{``We do not appreciate being experimented on'':
Developer and Researcher Views on the Ethics of Experiments on Open-Source Projects%
\thanks{\anon{[Funding anonymized]}{
Dror Feitelson holds the Berthold Badler chair in Computer Science.
This research was supported by the ISRAEL SCIENCE FOUNDATION (grant no.\ 832/18).}
}}
\author{\anon{\IEEEauthorblockN{[Authors anonymized]}}{
\IEEEauthorblockN{Dror G. Feitelson}\\
\IEEEauthorblockA{Department of Computer Science\\
The Hebrew University of Jerusalem, 91904 Jerusalem, Israel}
}\vspace*{-5mm}}

\usepackage{graphicx}
\usepackage{amsmath}
\usepackage{setspace}
\usepackage{graphicx}
\usepackage{adjustbox}
\usepackage[colorlinks=true, allcolors=blue]{hyperref}
\usepackage{pgfplots}
\usepackage{tikz,amsmath,amssymb,bm,color}
\usetikzlibrary{circuits.logic.US,circuits.logic.IEC,fit}
\usepackage{longtable}
\usepackage{array}

\usepackage{latexsym}

\begin{document}





\maketitle

\begin{abstract}
A tenet of open source software development is to accept contributions from users-developers (typically after appropriate vetting).
But should this also include interventions done as part of research on open source development?
Following an incident in which buggy code was submitted to the Linux kernel to see whether it would be caught, we conduct a survey among open source developers and empirical software engineering researchers to see what behaviors they think are acceptable.
This covers two main issues: the use of publicly accessible information, and conducting active experimentation.
The survey had 224 respondents.
The results indicate that open-source developers are largely open to research, provided it is done transparently.
In other words, many would agree to experiments on open-source projects if the subjects were notified and provided informed consent, and in special cases also if only the project leaders agree.
While researchers generally hold similar opinions, they sometimes fail to appreciate certain nuances that are important to developers.
Examples include observing license restrictions on publishing open-source code and safeguarding the code.
Conversely, researchers seem to be more concerned than developers about privacy issues. 
Based on these results, it is recommended that open source repositories and projects address use for research in their access guidelines, and that researchers take care to ask permission also when not formally required to do so.
We note too that the open source community wants to be heard, so professional societies and IRBs should consult with them when formulating ethics codes.
\end{abstract}
\begin{IEEEkeywords}
Experiments; Ethics; Open source
\end{IEEEkeywords}
\hide{

\ccsdesc[500]{General and reference~Experimentation}
\ccsdesc[500]{Social and professional topics~Codes of ethics}
\ccsdesc[500]{Software and its engineering~Open source model}
}
\begin{flushright}
\emph{That which is hateful to you do not do to another}\\
-- Hillel the Elder, Talmud Bavli, Shabbat 31:a
\end{flushright}

\section{Introduction}

The advent of open-source software development, and moreover the creation of repositories where numerous open source projects are hosted, has been a boon to empirical software engineering research.
Large volumes of high-quality code and related artifacts have become accessible for analysis \cite{harrison01,vonkrogh06}.
The next step was to use the record of changes to the code to also study the software development process.
A third step was to not only observe the code and the process, but to also interact with them.
For example, many research projects aimed at building tools for code improvement report on applying their tools on open source projects, and then submitting the improvement suggestions to the projects' developers.
The fraction of suggestions that is accepted is then used as an indication regarding the efficacy of the tool (e.g.\ \cite{tartler12,liuc13,monperrus19}).
While this practice exploits the time of the developers to assess the tool, it is considered acceptable as it conforms with the underlying principles of being ``open source'', which specifically include openness to contributions from anyone.

But in April 2021 the maintainer of the Linux kernel, Greg Kroah-Hartman, banned the University of Minnesota (UMN) from making contributions to the Linux kernel (for a detailed description of the whole affair, see e.g.\ \cite{chin21}).
He also reverted previous contributions from the university, pending a check that they are valid contributions.
The reason was a research project from the Lu lab in the university, dubbed the ``hypocrite commits'' (HC) study.
In this study patches that included bugs were intentionally submitted to kernel developers to see whether they would be accepted, under the pretext of demonstrating that open source development is vulnerable to malicious contributors \cite{wu:linux}.
In the email announcing the ban, Kroah-Hartman wrote ``Our community does not appreciate being experimented on''%
\footnote{\codefn{lore.kernel.org/lkml/YH\%2FfM\%2FTsbmcZzwnX@kroah.com/}}.

While many find this study clearly unethical, there are many other situations that are not so clear cut.
So when exactly do such studies constitute an unacceptable experiment?
This is not an easy question to answer, and there are myriad considerations including how exactly you define ``human subjects research'' and in fact whether the discussion should be limited to only ``research'' (e.g.\ \cite{watts17}).
Rather than theorizing on the definitions and the ethics considerations involved, we decided to elicit opinions straight from the horse's mouth.
We therefore conducted a survey among open source developers and asked about their reaction to various experimental scenarios.
For comparison, we also sent the survey to researchers involved in empirical software engineering research, and specifically those who use open source projects and experimentation.
As far as we know such a survey was never conducted among practitioners, and the last time a survey concerning ethics in computer science was conducted among researchers (or rather, university department heads) was 20 years ago \cite{hall01}.

The goal of the survey was to answer two main research questions:
\begin{enumerate}
    \item What do developers care about in terms of ethics in research?
    \item To what degree do researchers appreciate what developers care about?
\end{enumerate}
The survey started out by asking whether respondents knew of the Linux-UMN incident.
Subsequent questions asked about 16 possible concerns, ranging from looking at code without asking for permission to voicing an opinion about the quality of the work of identified developers.
The next section outlined 16 development and research scenarios, and asked respondents to judge the degree to which they were acceptable.
We collected responses from 168 developers and 56 researchers.
The response rate was reasonable for this type of survey:
about 9\% for the developers and 30\% for the researchers.
This attests to an awareness of and interest in ethics issues in such research, but also raises the danger of a selection bias, where the respondents are predominantly those with higher awareness and stronger opinions.

The results indicate that developers are quite open to various types of activity, be it novices who want to build a reputation, students who want to learn about open source, or researchers who want to study open source projects.
However, they also tend to expect that freedoms associated with the software and project be respected --- both their freedom to choose whether to participate in experiments, and maintaining the freedom of the code in line with its license.
Researchers largely see things eye to eye with developers, but sometimes do not fully appreciate the nuances of developers' opinions.
It is therefore recommended that repositories and projects explicitly address research use in their access guidelines, and that researchers approach them to ask for permission even if not formally required to do so.

The rest of the paper is structured as follows.
The next section provides background on ethics in research, starting with general ethics considerations and continuing with how they are applied to on-line research and to software engineering research.
Section \ref{sect:survey} then describes the survey and how participants were recruited.
Section \ref{sect:results} presents the results of the survey, followed by a discussion and recommendations in Section \ref{sect:disc}.
Finally, Section \ref{sect:threats} lists threats to validity, and Section \ref{sect:conc} presents the conclusions.

\section{Ethical Perspectives in Software\\ Engineering Research}

Experimental research on open source projects exposes some ethics considerations that have not been explored in the literature on research ethics.

\subsection{Background on Ethics in Research}

Experiments involving software developers fall under the regulations for ethical research involving human subjects.
The regulations covering such research were developed primarily in the context of biomedical research \cite{belmont}.
While each country naturally has its own precise regulations, the most commonly cited are those of the U.S.\ Department of Health and Human Services known as ``45 CFR 46'' (Code of Federal Regulations, Title 45, Part 46) \cite{hhs45cfr46}, which apply to research conducted or supported by U.S. government agencies.
These regulations define research on human subjects as ``systematic investigation, including research development, testing, and evaluation, designed to develop or contribute to generalizable knowledge'', where ``an investigator (whether professional or student) conducting research: (i) Obtains information or biospecimens through intervention or interaction with the individual, and uses, studies, or analyzes the information or biospecimens; or (ii) Obtains, uses, studies, analyzes, or generates identifiable private information or identifiable biospecimens.''

Using the HC study as an example, Wu and Lu wrote in their paper that ``The IRB%
\footnote{Institutional Review Board, the committee charged with reviewing research proposals to ensure they are ethical.}
of University of Minnesota reviewed the procedures of the experiment and determined that this is not human research. We obtained a formal IRB-exempt letter'' \cite{wu:linux}.
More specifically, in a FAQ related to the paper they wrote ``This is not considered human research. [...] We send the emails to the Linux community and seek community feedback'' \cite{wu:linux-faq}.
However, a community is composed of individuals, and in the end they did indeed interact with individuals and obtained information about their behavior.
Moreover, they themselves note the need to protect the identity of the Linux maintainers who handled their patches, as being identified as having approved faulty code may cause embarrassment or other inconvenience \cite{wu:linux}.
In addition, they acknowledge the problem of wasting maintainers' time, which was the more important issue for at least some of the Linux maintainers.
It seems that the ethics of a scenario of observing people perform their work, and adding to that work as part of the experiment, has not been specifically discussed in the literature.

The basic document on the ethical use of human subjects in research is the Belmont Report from 1979 \cite{belmont}.
The Report first makes a distinction between research and normal practice, and states that if an activity contains any element of research, it should undergo an ethics review.
It then identifies three basic ethics principles:
\begin{enumerate}
    \item \textbf{Respect for Persons}, so experimental subjects should be informed about the experiment and are entitled to decide for themselves whether to participate in it;
    \item \textbf{Beneficence}, comprised of the obligations to do no harm and to maximize possible benefits; and
    \item \textbf{Justice}, meaning that the benefits of the research as well as its costs should be shared equally by all.
\end{enumerate}
The HC study clearly fails the first, in that informed consent was not sought.
Moreover, a central tenet of open source development is trust, based on reputation and a hacker ethic of sharing \cite{delaat10,delaat14}.
The report on the incident by the Linux Foundation Technical Advisory Board specifically cites the breach of trust that contributions to the kernel are well-intended as a major offense, and in fact classifies the whole affair as a ``breach-of-trust incident'' in its title \cite{linux-umn-rep}.
One can also argue that the HC study fails the second principle, as it wasted maintainers time.
This was aggravated by the trust problem, as considerable work was needed to re-assess all previous UMN contributions that had been reverted \cite{linux-umn-rep}.
The study does seem to comply with the third principle, in that the maintainers who's time was wasted were not singled out for any particular reason.

Professional societies publish ethics codes for their members which often mention the Belmont Report.
One example is The American Psychology Association Ethical Principles of Psychologists and Code of Conduct.
While the bulk of this code concerns the professional conduct of psychologists, it also includes a section about research, with a detailed description of what should be included in informed consent, and how to deal with cases where deception is necessary to elicit spontaneous behavior \cite{apa:ethics}.

The IEEE Code of Ethics, in contradistinction, focuses exclusively on professional conduct and does not mention research at all \cite{ieee:ethics}.
The ACM Code of Ethics and Professional Conduct also does not address the issue of ethics in research, except in saying that ``The public good should always be an explicit consideration when evaluating tasks associated with research'' and many other activities \cite{acm:ethics}.
Informed consent is mentioned only in relation to respecting privacy: as a computing professional, if you build a system that collects data, you should ensure ``individuals understand what data is being collected and how it is being used''.
This is also the case for the IFIP Code of Ethics and Professional Conduct, which is based on the ACM code \cite{ifip:code}.
However, the ACM recently approved a new ``policy on research involving human participants and subjects'', which specifically includes minimization of potential harm and adherence to informed consent and justice \cite{acm:human}.

\subsection{Ethics in Online Settings}

Mining repositories of open source software, and in particular collecting data on software developers, is just a special case of collecting social-science data about the behavior of individuals in some context.
In some cases this has clear ethical implications, for example when studying the consumption of porn \cite{thomas96b}, postings in support groups for trauma victims \cite{king96}, or an LGBT forum including discussions of coming out \cite{bassett02}.
\hide{
Thomas summarizes the basic ethics guidelines in cyberspace thus \cite{thomas96b}:
\begin{enumerate}
    \item Never deceive subjects;
    \item Never knowingly put subjects at risk; and
    \item Maximize public and private good while minimizing harm
\end{enumerate}
}
But even if there are no clear repercussions to divulging data about participants, there is still the question of the expectation for privacy and the boundary between ``public'' and ``private'' in an online setting \cite{thomas96a,bassett02}.
Such expectations exist despite Google's search capabilities, which often allow the identification of even short quotations from text (or code) which is openly accessible \cite{oezbek08,gold22}.

As the above examples imply, online ethics often apply to the \emph{users} of software systems, especially web-based systems.
Facebook in particular provides several relevant case studies with increasing severity.
The first level is using A/B testing to improve the product \cite{feitelson13}.
This means that new features are added only after being tested with real users in live action, unbeknownst to those users, to see that they lead to a favorable response.
This practice is widely used in practically all web-based companies \cite{kohavi09a,fabijan18,li21}, and Google even offers it as a free service on its marketing platform
\footnote{\codefn{marketingplatform.google.com/about/optimize}}.
However, it introduces two ethical difficulties.
First, there is no informed consent by the users to participate in the experiment.
This may be excused on the grounds that users don't know what version of the system they are using anyway, and this changes daily.
But this leaves the second issue, which is what the experiment is really trying to optimize.
It may be that the benefits are for the company and not for the users, e.g.\ when the desired effect is to improve the monetization from using the product, even if this comes as the expense of the users in terms of expenses (on ecommerce sites) or detrimental effects like addiction.

The second level is facilitating other research, which is not directly related to improving the service to users \cite{king15}.
For example, a much-cited research done using Facebook data concerned the transmission of emotions among social network users \cite{kramer14}.
To do so, ``the experiment manipulated the extent to which people (N = 689,003) were exposed to emotional expressions in their News Feed''.
This manipulation was justified by Facebook being a private company, hence not bound by regulations on research supported by the government, and by the fact that the analysis performed ``was consistent with Facebook’s Data Use Policy, to which all users agree prior to creating an account on Facebook, constituting informed consent for this research''.
However, this lenient interpretation of informed consent elicited an expression of concern from the Editor-in-Chief of the journal where the research was published \cite{verma:pnas-ed}.

The third level is facilitating direct manipulation of the public, as in the infamous Cambridge Analytica affair (for a detailed description of this affair, see e.g.\ \cite{meredith18}).
On a superficial level the ethical problem was that around 270,000 Facebook users gave informed consent to use their Facebook data for psychological profiling, but using Facebook's Open Graph platform the profiling app actually collected data also from their Facebook friends' accounts, totaling 87 million users.
The more insidious problem was that Cambridge Analytica claimed to have used this data to manipulate voters in the 2016 USA elections, which may have contributed to the campaigns of Senator Ted Cruz and President Donald Trump.

\subsection{Application to Software engineering Research}

Vinson and Singer attempt to adapt the Belmont principles for the special case of experiments in software engineering \cite{vinson08,singer02}.
They retain the first two principles, of informed consent and beneficence, and add two more:
maintaining the \textbf{confidentiality} of all information shared by the experimental subjects, and ensuring \textbf{scientific value}, in particular by using established methodology.

\newcommand{\vv}{\rule{0pt}{2.5ex}}
\begin{table*}\centering
    \caption{\label{tab:considerations}\sl
    Potential ethics issues in studies using open source projects.}
    \begin{tabular}{p{0.35\textwidth}p{0.5\textwidth}}
    \hline
    \emph{Study type} & \emph{Potential issues} \\
    \hline
    Studying the code
        & Use of the code not for the purpose for which it was opened \\
    \vv & Harming a project by judging it or publishing its faults \\
    \vv Studying the developer community
        & Violating developers' privacy \\
    \vv & Harming developers by exposing inappropriate practices \\
    \vv Interacting with a project's development
        & Harming the project by adding inappropriate code \\
    \vv & Wasting the time of the projects' original developers \\
    \hline
    \end{tabular}
\end{table*}

The Menlo Report on Ethical Principles Guiding Information and Communication Technology Research is also based on the Belmont Report.
It adopts the three original principles, and adds \textbf{Respect for Law and Public Interest} \cite{menlo}.
However, in doing so the Menlo Report actually conflates ethics with legal issues and with safety.
This is a result of two factors.
First, their focus is on security research, including whitehat hacking and the study of malware.
Studying malware can be dangerous, but this is not an ethical issue but a safety issue, just like studying explosives in a chemistry lab or viruses in a biology lab are not ethical issues but safety issues.
Second, the law may indeed step in when ethics standards fail.
For example, the European GDPR and the California CPA were both responses to lack of respect for privacy and information security on the side of high-tech companies.
But while this may affect research, the motivation was unrelated to research.
And ethics is more than just abiding by the law.
In fact, ethics guidelines are specifically an attempt to codify what is right or wrong \emph{beyond} what the law demands.
So the Menlo Report is not so much about the ethics of human subjects research, as about all aspects of research with the \textbf{potential to harm humans}, mainly by exposing data about humans.

Experimenting with software engineers can lead to various ethical issues, including inconvenience due to frustration or boredom during the experiment, worrying about it, and disapproval or stigmatization by co-workers due to disclosed information \cite{sieber01b}.
There is also a growing body of software engineering research that interacts with the subjects directly, e.g.\ using fMRI or psychological tests \cite{floyd17,peitek20,graziotin22}.
These are adequately covered by procedures used in other fields which use such devices, e.g.\ cognitive science.

Additional vexing ethical questions may come up in the context of collaborations between researchers and industry \cite{lethbridge01}.
For example, consider a company-based study where developers are ranked based on a quality analysis of their code.
Should the researchers be loyal to the company, which invited them to collaborate and may be financing them, and would be harmed by keeping sub-par employees --- or to employee-subjects, who trust them with their data, and might be fired? \cite{vinson01}.
As it is hard to anticipate and regulate all such potential conflict situations, they imply a need for open discussions among the collaborators to map out the issues and decide on how to deal with them.

A recurring subject in previous investigations on ethics in software engineering research is the use of student subjects in experiments.
The risk here is that the researchers are also the professors, and the situation might be perceived by the students as coercive \cite{liebel21,singer02,sieber01b}.
Thus at a minimum one needs to uphold anonymity, and allow the option to opt out, thereby negating the fear of influence on grades.

There has also been concern regarding harm to the students' academic progress.
Therefore, especially when experiments are carried out as part of compulsory classes, they should have educational goals \cite{singer02,carver03,berander04,carver10}.
Examples include the opportunity to learn or exercise some technique or methodology, being exposed to cutting-edge ideas and procedures, and more \cite{staron07}.
At the same time, one should consider whether the students could have learned the same things more efficiently in some other way, and whether it is fair to grade them on their performance in an experiment, especially if they were divided into groups that used different treatments?

\subsection{Considerations for Open Source}

Most of the research using open source projects is based on repository mining.
In addition to the repositories themselves, Internet sites like OpenHub%
\footnote{\codefn{www.openhub.net}.}
are devoted to the tabulation and display of the activity by different developers in different projects, adding to their exposure.
While often thought to be benign, such exposure can in fact have detrimental implications (Table \ref{tab:considerations}) \cite{gold22}. 
For example, OpenHub ranks developers by ``kudos'', which is assigned by members of the site to each other.
Research may identify ``influential'' or ``core'' developers, using measures of impact usually based on levels of activity (e.g.\ \cite{wu18,yan20,song22}).
Such rankings may shame or offend those who receive a low ranking.
Even more harm can ensue if developers or companies are ranked based on a quality analysis of their code, or if intellectual property is revealed \cite{vinson01}.

Open source projects are also vulnerable to situations where researchers interact with the developers, as in the HC study.
The goal of the HC study (which led to the ban on UMN contributions to the Linux kernel) was to raise awareness of how vulnerabilities can be introduced to open-source software on purpose by malicious agents \cite{wu:linux}.
The idea was to analyze the code and create ``hypocrite'' patches that added a missing condition for a vulnerability, thus turning ``immature vulnerabilities'' into real vulnerabilities.
As a proof-of-concept, they prepared 3 patches that created ``use after free'' bugs in Linux%
\footnote{In total it appears that 5 patches were submitted, but only 3 were buggy \cite{linux-umn-rep}.}.
They claimed that this was done safely, and that the buggy code was only exchanged in emails,
and indeed the 3 buggy patches were rejected by Linux maintainers.
But the main fault developers found with the study was the lack of obtaining consent.

Other unique ethics issues with open source concern whether and how the research interacts with the open source philosophy.
At the most basic level the issue is one of openness and freedom.
On the face of it there should not be any problem.
A well-known adage regarding free software likens it to free speech as opposed to free beer.
The implication is that anyone can do whatever they want with the code \cite{chopra09,wolf09}, and the problem is not in exposing it but only when it is confined and restricted.
Indeed, the hacker ethic of free information justifies using systems in unintended ways to uncover their inner working \cite{chopra09}.
However, one may question whether it is ethical to use public data for purposes other than intended by its authors \cite{elemam01}.
Opening source code is not done to facilitate research --- it is done to improve the code and benefit its users.
So using it for other purposes may require the consent of the authors.
But who do you ask for consent for using code from a long lived project where developers come and go with time? \cite{gold22}

The flip side of openness is the danger of compromising privacy \cite{zimmer10}.
Vinson and Singer comment that developers who identify themselves as authors of open-source code cannot expect privacy, and therefore using their code --- and even identifying them --- does not fall under the usual limitations of human subjects research \cite{vinson08,singer02}.
But according to Berry, the question is ``Is the Internet a space in which embodied human beings interact? Or is it a textual repository where authors deposit work?'' \cite{berry04}.
The difference brings up considerations like copyright and fair use of artifacts, not necessarily from the legal perspective, but from the social perspective.
For example, software published under the GNU public license requires any derivative work to be published under the same open license.
Legally speaking, quoting from such software in research probably falls under fair use.
But ethically, is it acceptable to republish even just short pieces of code in copyrighted papers, given the specific license provisions? \cite{berry04}.

Moving beyond privacy, removing restrictions on using open source data may lead to risks that developers may not be aware of.
One example of the need to keep information confidential is when studying the development process, and some employees do not follow prescribed practices \cite{beckerk01};
exposing this is part of the research goal, but if the employees are identified they might be punished or even fired.
As this example shows the risk also depends on the style of the research.
There is a difference between actual reading and analysis by human researchers and the publication of specific identified quotes, and massive-scale automated analysis using machine learning, where the end result is just statistical observations.
And indeed, developers are sometimes sensitive to their privacy, and even may disassociate themselves from code they had written, as evidenced by the existence of services like Gitmask%
\footnote{A project which facilitates anonymous contributions to open-source projects, \codefn{www.gitmask.com}}.
This casts a shadow on research practices such as analyzing developers' emails \cite{bacchelli10}.

A more philosophical level concerns the issue of contribution.
At the core of open source software development lies the notion that anyone is invited to contribute to the project.
So the basic expectation is that people will give to the project, especially if they also benefit from it.
For example, Grodzinsky et al.\ write about the ethical responsibilities of open source developers, including the obligation of organizations who use open source software to also contribute to the community, and the obligation to produce high-quality code \cite{grodzinsky03}.
Yu writes about the reciprocity in firms' open source policies, and how it contributes to their business performance \cite{yu20}.

Finally, it is not clear that a legally or even ethically centered discussion is the right approach.
As Bakardjieva and Feenberg write,
``alienation, not privacy, is the actual core of the ethical problems of virtual community research'' \cite{bakardjieva00}.
Our survey is specifically designed to obtain input from the community itself about what really concerns them.

\section{Survey Design and Execution}
\label{sect:survey}

To the best of our knowledge the closest work to ours is a superficial survey of department heads regarding awareness and procedures for ethics approval of software engineering experiments with human subjects, published by Hall and Flynn twenty years ago \cite{hall01}.
We conducted a deeper survey of the considerations involved, as seen by the researchers themselves, and, more importantly, by the potential experimental subjects.

\subsection{Survey Structure}

The survey contained four sections:
\begin{enumerate}
    \item Questions about the Linux-UMN incident;
    \item Questions on general ethical considerations when performing research on open-source projects;
    \item A list of short scenarios describing contributions to open-source projects or research on such projects, asking for judgment on whether they constitute acceptable behavior;
    \item Demographic questions used to characterize and classify the respondents.
\end{enumerate}
The questions are detailed below together with the results.
Each section ended with an option to provide general comments.

In writing the questions about ethics considerations, we used concrete questions rather than asking about abstract principles.
For example, instead of asking about the principle of not putting experimental subjects at risk, we asked about voicing an opinion about the quality of developers' work.
Likewise, instead of asking about the principle of maintaining privacy we asked about reading and analyzing the text of communications between developers to better understand their social interactions.
The respondents were asked to rank how much these actions are an ethical concern on a 7-point scale.

The questions on acceptable behavior also used concrete scenarios that may happen in the work on an open-source project.
Some of the scenarios focused on developers, for example a developer who contributes code that was not adequately tested.
Other scenarios were about researchers, for example identifying a project whose development they had analyzed.
The respondents were again asked to judge whether they are acceptable behaviors on a 7-point scale.

The survey was implemented on the Google forms platform.
None of the questions were mandatory, and no identifying information was collected.

\subsection{Recruiting Subjects}

The recruiting procedure and its outcome are summarized in Table \ref{tab:recruit}.
Given the nature of the topic we aimed to collect the opinions of more experienced developers and researchers, rather than novices and students.
To ensure we had subjects of both types we employed separate procedures to recruit developers and researchers.
However, the distinction is not really binary, as researchers may also make code contributions to open source projects, and developers may participate in research.
We therefore also asked the subjects about how they identify themselves, and this was used to adjust the final classification as described below.

For developers we wanted those who were involved with the project and not just making a casual contribution.
We first selected active recent GitHub projects, identified by a threshold of having at least 50 commits since 2020.
To ensure that they are software projects we used the CCP (corrective commit probability) metric, and used only projects where this was between 0 and 1 \cite{amit21}.
This excludes projects which do not have commits that are identified as bug fixes.
From these projects we extracted developers who had public emails and at least 20 commits since 2020.
There were 16,559 such developers.
From them we randomly selected 2000 and invited them to participate in the survey.
The invitations were made by personal emails.
In most cases these were addressed using the first name, after manual checking.
When the first name could not be identified the email was addressed to ``developer''.
17 emails bounced, and 180 responded, leading to a response rate of just over 9\%.
This is significantly higher than the 2--4\% reported by Wagner et al.\ \cite{wagner20}.

\begin{table}\centering
    \caption{\label{tab:recruit}\sl
    Summary of recruited participants.}
    \begin{tabular}{@{}lc@{~~}c@{~~}c@{~}c@{}}
    \emph{Inclusion criteria} & \rotatebox{75}{\emph{Invited}} & \rotatebox{75}{\emph{Bounced}} & \rotatebox{75}{\emph{Responded}} & \rotatebox{75}{\emph{Resp.\ rate}} \\
    \hline
    \multicolumn{4}{c}{Developers} \\
    \hline
    $\ge 20$ commits since 2020  & 2000 & 17 & 180 &  9.08\% \\
    in project with $\ge 50$ commits   & & & \\
    \hline
    \multicolumn{4}{c}{Researchers} \\
    \hline
    published on experiments or  &  151 &  3 &  44 & 29.7\% \\
     open source; H-index $\ge 10$    & & & \\
    \hline
    \end{tabular}
\end{table}

To identify relevant researchers we performed a Google Scholar search with the query
``\textsf{\small "software development" experiment "open source" github}''.
The query was restricted to papers published in the last 5 years (since 2016).
We then verified that the paper is indeed on topic based on its title (there were a few irrelevant ones, e.g.\ reporting on open-source software developed to analyze a physics experiment) and published in a leading software engineering venue.
From these papers we selected authors who have an H-index of 10 or above.
The first 150 papers returned by the query yielded 48 usable papers and 102 authors.
Some of the papers were not used because all their above-threshold authors had already been identified from previous papers.
The authors were again invited to participate using personal emails, addressed to their first names.

To increase the number of research participants, we conducted a second wave of invitations, based on the ``empirical software engineering'' label which researchers can attach to their Google scholar profile.
The top 100 researchers with this label were scanned, and those with papers with titles indicating work on experiments and open source in either the first page or the last 5 years were identified.
There were 57 such researchers, but 8 of them had already been identified in the previous round, so only 49 additional invitations were sent.
In total then 151 invitations were sent, of which 3 bounced.
44 authors responded, leading to a response rate of nearly 30\%.

Participants in the survey were not paid or given any other reward.
The instructions indicated that advancing to the questions constitutes consent to participate.
The survey and recruitment procedure were submitted to the ethics committee for non-medical research on human subjects of the faculty of science, and received approval.
Due to rate limitations on sending emails, the invitations were sent over several weeks from the end of October to the end of November 2021, about 6 months after the Linux-UMN incident.

We are aware of the problems with sending unsolicited emails to invite potential participants to a survey \cite{baltes16,cho99}.
The above procedure was meant to reduce the danger of sending such emails to irrelevant people, and the relatively high response rates indicate that indeed many recipients found the survey relevant and important.
Some even said so explicitly and expressed interest in the results.
This issue is discussed further below, in the context of a survey question that addressed it and in the recommendations.

\subsection{The Survey Respondents}

A total of 180 active GitHub users responded to the developers survey, and 44 Google Scholar authors to the researchers survey.
However, some of the developers identified themselves as also being researchers and reported having authored multiple papers on empirical software engineering.
And some of the researchers reported significant activity in open-source development.
We eventually reclassified 12 respondents to the developers survey as researchers, as they had self identified as researchers and not as paid developers.
Subjects who reported being both researchers and developers were left in their original classification.
All the results presented below are after this reclassification.

\input{plt-43}

Interestingly, the distribution of years of development experience was similar for developers and researchers, with developers having only a slight edge.
However there was a large difference in the distribution of number of papers published:
78\% of developers had published none, and the maximum was 10.
Only 14\% of researchers had not published papers (or did not reply), and the median was 20.
The distributions are shown as CDFs (cumulative distribution functions).
This enables an easy comparison of the distribution of responses of developers and researchers.
A CDF that is below and to the right of another implies a distribution biased towards higher values.

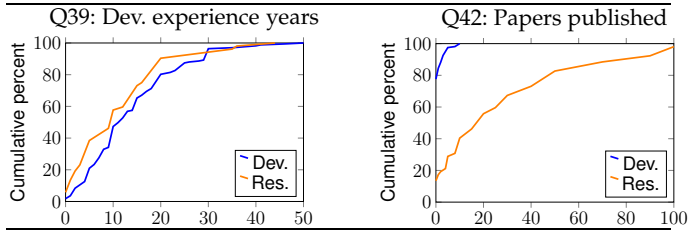
\input{plt-39}

Regarding experience with open source projects, more than 30\% of researchers reported having none.
And the distributions of years of experience and number of projects worked on by developers dominated the respective distributions of researchers.

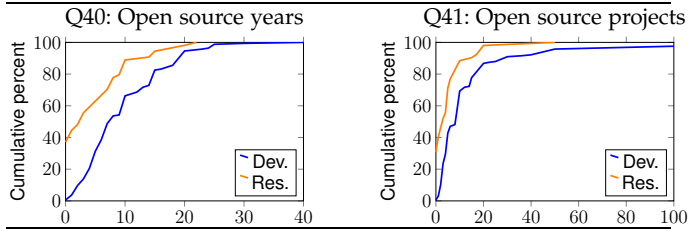
\input{plt-40}

\section{Survey Results}
\label{sect:results}

The results for the survey questions are shown as histograms of the selected options.
In most questions the options form a scale.
In these cases a CDF is shown too.

\subsection{The Linux-UMN Incident}

The first part of the survey concerned the Linux-UMN incident.
The first question was whether the survey participants had heard about this incident.
As shown below, a substantive majority had heard about it, and most --- and even slightly more so among developers --- had followed it when it happened.
In other words, this incident was an important news story for our participants.

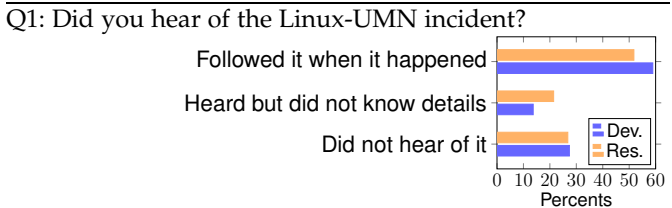
\input{plt-1}

Delving into the details, the majority of respondents thought it was justified to ban UMN from contributing to the Linux kernel.
Nearly all the rest thought this response was somewhat exaggerated%
\footnote{This and some of the options in other questions are shown in abbreviated form to fit in the available space.},
and only a handful thought it was wrong.
In added comments, quite a few respondents referred to the gap between the open source community and academic researchers, some even using quite strong language concerning researchers in their ivory towers.
One wrote: ``If you wouldn't perform the same experiment on commercial developers without management consent, or on academic colleagues without university consent, and you didn't get permission from project owners, don't do it''.
Particular ire was reserved for the perception that researchers were using developers but then were not interested in hearing their opinions and did not solicit feedback.

\input{plt-2}

Responses were more evenly distributed concerning the question whether the UMN IRB had erred in determining that this is not human research.
Around 13\% of developers and researchers thought they got it right.
The rest of the developers were evenly split between claiming they were wrong and saying it is hard to tell; among researchers a significant majority said they were wrong.

\input{plt-3}

When asked to identify the worst offense in the UMN study, the top spots were wasting the time of maintainers (preferred by researchers) and the risk of distributing buggy code (preferred by developers).
Some explicitly cited the lack of informed consent and violating trust; in the graph these are included under ``treating the Linux developers as guinea pigs''.
Others said all 3 offenses were equally bad; they were counted as contributing $\frac{1}{3}$ to each.

\input{plt-4}

Several comments made at the end of the survey can be interpreted as alluding to this question.
Some referred to the special status of Linux as a major global resource, so any interference with it is especially troubling.
Another interesting comment was: ``The biggest sin of the research was that it was done without cooperation with the community''.
The relationship between academic researchers and open source developers figured in many other comments as well, and is discussed in Section \ref{sect:disc}.

It is interesting to note that hardly anyone thought the study was OK.
This does not contradict the previous question, where around 13\% said it was not human research at all, because of the option to find fault with the study for reasons other than ethics --- and specifically because of the danger that buggy code would enter the distribution.

Finally, there was a wide range of opinions on whether this study could have been executed in an ethical manner.
Specific interesting results were that many researchers and not few developers thought it would be OK only if informed consent was given, even though such consent may harm the validity of the experiment.
On the other hand many developers were also content with having the experiment cleared with project leaders without explicit informed consent from the affected maintainers.

\input{plt-5}

In comments one developer cited the practices of whitehat hackers to contact the security-focused maintainers of a project to coordinate the scope of the research up front.
Others claimed the results could be obtained by other means.
In addition, several developers suggested that if maintainers' time was wasted by an experiment they should be compensated, either directly or by a donation to the project.

\subsection{Ethics Concerns}

The second part of the survey was about possible ethics concerns in isolation.
The respondents were asked to rate these concerns on a 7-point scale, ranging from 1 = no concern to 7 = extreme concern.
This directly reflects our first research question, of what developers care about.
The results also pertain to the second research question by comparing the answers of developers to those of researchers.

We present the results in separate subsections that each contains a set of questions related  to the same general concern.
Note that in the survey there was no such structure, and in some cases the questions appeared in a different order.
The original order is given by the question numbers.

\subsubsection{Inappropriate use of open source code}

The first possible concern is about using open source code in a manner different from the intentions of its developers.
One question asked about using code examples without asking permission.
A large majority of both developers and researchers thought this was of little or no concern.
The second question was about quoting code in copyrighted articles.
In this case developers were consistently more concerned than researchers.
Quite a few added comments about open source licenses prohibiting this, as they require all derivatives of the code to remain free.
Others commented that in general one should learn about a project's rules and culture before using it in research, and respect these rules.

\input{plt-6}

\subsubsection{Expectations for privacy and confidentiality}

Several questions concerned the exposure of the developers.
A set of three questions were about using the texts written by developers to communicate among themselves.
Developers were very open to having such texts read when the goal was to better understand technical issues;
researchers were slightly more reserved.
Both developers and researchers were somewhat more reserved when the goal was to understand social interactions.

\input{plt-8}

Interestingly, developers were even more reserved about the use of machine learning to analyze all the communications and derive a statistical characterization.
One explained in a comment that machine learning may not catch all nuances of human communication and especially cases of non-native-English speakers or jokes.

\input{plt-10}

A second set of questions concerned the identification of developers in research reports.
These questions elicited a wide range of responses, from those who saw no problem even with the version asking about researchers voicing an opinion about the quality of work of identified developers, to those who were gravely concerned with the version that just asked about identifying developers who had contributed to a project.
Tellingly, there was much less concern regarding voicing opinions on the project as a whole as opposed to identifying individual developers.

\input{plt-18}

In all these four questions, researchers were significantly more concerned than developers, perhaps due to recent increased awareness of ethics issues in research.
It could also be a matter of culture: one developer commented that in the open source culture criticism is welcome, but it should be a constructive discussion on how to improve and not a judgment after the fact.
Another wrote ``being a part of experiment or being judged by some non-even-a-contributor will probably lead to [...] ending any contributions''.

\subsubsection{Interfering with developers' work}

A related issue is interacting with developers in a way that might interfere with them.
There was wide agreement that asking developers about their work is of no or at most little concern.
However, the majority were opposed to engaging developers without explaining that this is part of an experiment and obtaining informed consent.
In both questions, researchers were marginally more concerned than developers.
A third question about deceiving developers so as not to affect their behavior elicited nearly uniform responses, with slightly higher concern by developers.

\input{plt-11}

The concern regarding interfering with developers may be modulated by the merits of experiments.
Questions about this led to a wide range of responses.
The majority were concerned about the possibility of experiments with no scientific value, but there was also a sizable minority who thought this was of no concern.
And there was a nearly uniform response to the possibility of using a new experimental methodology, albeit with more respondents expressing no concern than any other single option.

\input{plt-14}

\subsubsection{Risk of harming the project}

Finally, developers and researchers alike were very concerned about conduct that may cause harm to a project.
There were two questions about this, one asking about contributing buggy code and the other about wasting the time of maintainers to review and reject code.

\input{plt-16}

\subsection{Acceptance of Scenarios}

The next part of the survey presented several scenarios and asked whether they were acceptable.
A 7-point scale was used, as follows:
\begin{description}
\item [-3] criminal
\item [-2] unacceptable
\item [-1] preferably not done
\item [~0] I'm not sure
\item [~1] not great but tolerable
\item [~2] reasonable and acceptable
\item [~3] best practice
\end{description}
Respondents were asked to try and be categorical, using 2 and -2.
As in the previous section, we present the scenarios here in groups that do not necessarily correspond to the order in which they were presented in the survey.

\subsubsection{Contribution for self benefit}

Three of the questions were about contributing to an open-source project for self benefit, rather than to advance the project.
In essence this reflects the ``scratch a personal itch'' motivation identified by Raymond \cite{raymond:cab}.
The results show that this is an acceptable or at least tolerable practice, with only a small minority of respondents indicating that preferably it would not be done.

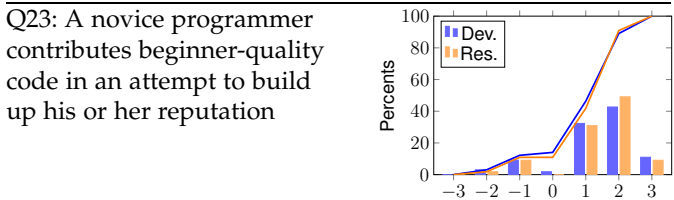
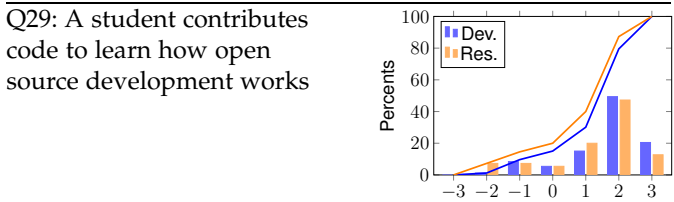
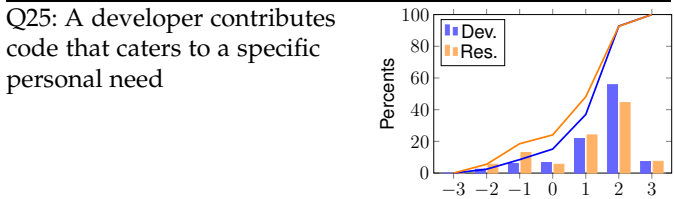
\input{plt-23}

A special case of contributions that are not made just to advance the project is when the contribution is part of research on open source development (like the HC study was).
There was wide acceptance of the use of open source projects to perform research about code and its development, with many even calling it a best practice.
At the same time there was strong opposition to the idea of submitting buggy code to see if it would be caught, and in the case of security bugs many called such behavior criminal.

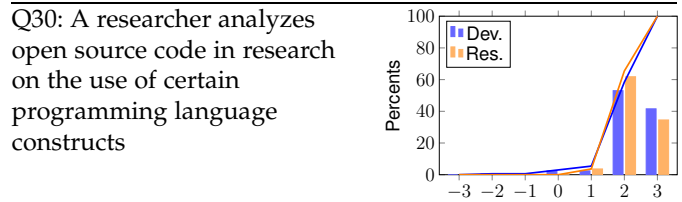
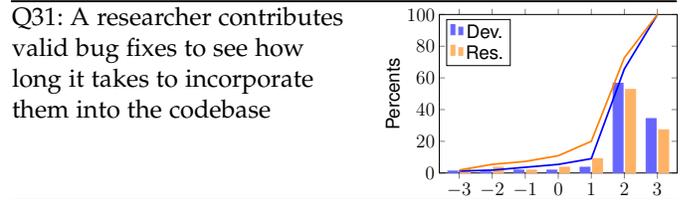
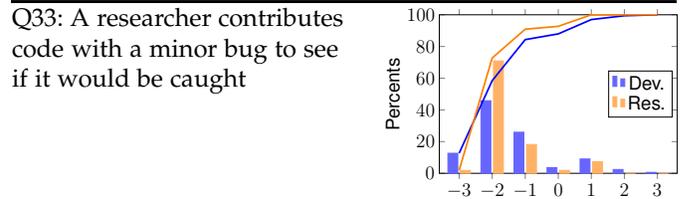
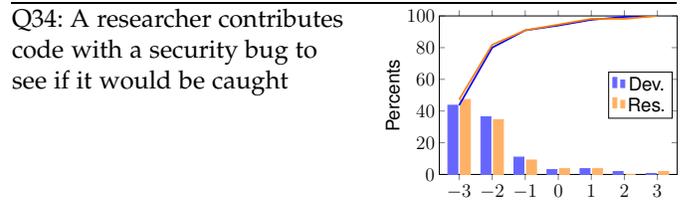
\input{plt-30}

\subsubsection{Contributing potentially problematic code}

One of the dangers with contributions made for self benefit was that the submitted code may be sub-par.
Making this explicit, a pair of questions concerned the role of testing.
The responses indicate that contributing code that was tested is acceptable even if it still contains a bug (which the tests failed to uncover), while contributing code that was not adequately tested was usually thought to be unacceptable or at least something that should not be done, although some also thought it was tolerable.

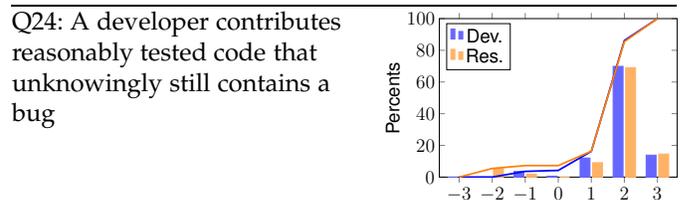
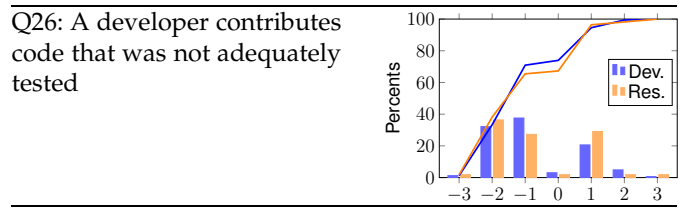
\input{plt-24}

Continuing this sequence, three questions concerned the contribution of code generated by an automatic tool.
Contributing such code after checking it manually was generally considered acceptable.
But contributing code from an experimental tool to check whether it would be accepted, thereby obtaining an evaluation of the tools quality, elicited a wide range of less favorable responses.
Interestingly, the responses were somewhat more accommodating for experimental tools by researchers than for novel tools by developers.
In comments, several respondents suggested that automatically generated patches should be identified as such.
This would run the risk that developers are prejudiced against tools \cite{monperrus19}.
And another comment was that only the code is really important, and not who or what produced it.

\input{plt-27}

\subsubsection{Identification of developers}

Reports of research on open source development may identify the project and developers that were used in the research.
Identifying the project was generally viewed as acceptable, but when the interactions among developers were studied, there were slightly more developers who thought that it should not be done.
Identifying developers who wrote specific commit messages met with significantly stronger opposition, especially from researchers.
This correlates with researchers being more sensitive to privacy as we saw above.

\input{plt-36}

A special case of identification is sending surveys to the emails of developers of open-source projects (like this survey was conducted).
This was generally considered acceptable or at least tolerable, although a non-negligible minority said that preferably it would not be done.
This result contradicts Baltes and Diehl, who quote a developer who said that such unsolicited surveys are ``worse than spam'' \cite{baltes16}.
And the present survey also received one response equating academic surveys to spam.
But based on our results it may be that only a small minority indeed view such practice as unacceptable.
However, if developers who are opposed to such invitations refrained from answering the survey (as the one we received said he does), this result is biased.
We have no way to know how many potential respondents actually think that sending unsolicited questionnaires is unacceptable, but then didn't register this opinion for this very reason.

\input{plt-35}

Wagner et al.\ report on a similar situation, where an addressee of a survey approached GitHub to check on them \cite{wagner20}.
The result was a determination that they had not violated GitHub's terms of service, but a suggestion to check beforehand in the future.
GitHub documentation indicates that current practice is to set new users' emails to ``private'' by default.

\section{Discussion and Recommendations}
\label{sect:disc}

From a perspective of over 40 years, and the vantage point of software engineering research, it seems that the Belmont report is perhaps not the best basis for discussing ethics.
The terminology of ``respect for persons'' and ``beneficence'' obscures some of the real concerns of practitioners.
Likewise, using US government definitions as a basis for discussions about what exactly constitutes ``human research'' is distracting and unproductive.

Open source developers in particular seem to want to promote good, which is what beneficence is all about.
This starts, of course, with the drive to create good and useful software, and making it free for anyone to use.
But it also includes giving others a chance to interact with the community and to develop professionally, as witnessed by the positive responses to our questions about a novice programmer contributing beginner-quality code, or a student wanting to learn how open-source works.
A relevant example is the response to one of the buggy patches submitted to the Linux kernel as part of the HC study, where the maintainer who handled it gave suggestions on what might be done to improve it%
\footnote{\codefn{lore.kernel.org/lkml/20200821081449.GI5493@kadam/}},
in an apparent attempt to mentor what appeared to be a not-very-proficient junior contributor.

This openness and willingness to help naturally extends to researchers.
It is OK to use code examples without asking permission.
It is acceptable and even a best practice to analyze the code and the project's history.
It is perfectly fine to read communications between developers, and even to approach them directly, in order to better understand the project.
There is also no problem with contributing valid code to the project to follow how it is treated.

At the same time, what developers care about is
\begin{itemize}
    \item \textbf{Maintain transparency}: If your goal is research and not contribution to the project, state this outright.
    Give details of your research, obtain informed consent, and take responsibility for your work.
    In general deception and masquerading are frowned upon;
    however, when justified by the nature of the research it may be acceptable to coordinate the level of disclosure with project leaders.
    \item \textbf{Do not harm the project}.
    In particular do not harm the code or put it at risk.
    The whole ethos of open source development is to improve the code;
    doing the opposite puts you on a collision course with the community.
    \item \textbf{Do not clash with the developers and maintainers}.
    Many of a project's developers and maintainers are volunteers, and their time and good will are the scarcest and most valuable resources at the disposal of the project.
    Negative interactions may lead to frustration and reduced willingness to contribute.
    In particular, listen to the developers and do not pass judgment on the project or on the activity without giving them a chance to explain themselves.
    The goal should be to learn and improve, never to blame.
    \item \textbf{Follow the rules}, such as license restrictions%
    \footnote{Regrettably this is not as simple as it sounds as there are so many different variants \cite{stlaurent04,zacchiroli22}.}
    and customs.
    Open source development is a community.
    If you want to participate, you need to accept the community standards.
    For example, the fair use argument for publishing code excerpts may be irrelevant, because the developers do not see this as a legal issue but rather as a core values issue.
\end{itemize}
Upon reflection, these considerations can be generalized and summarized as requirements for \textbf{maintaining and justifying trust} in the good intentions of the researchers.
One respondent explicitly wrote in a comment that ``Some actions may be worth forbidding even if they are not immoral because they damage the reputation of science as an institution and limit future opportunities for cooperation''.
Trust can be further strengthened if the researchers reciprocate and contribute to the projects they study.

One issue which suffers from a significant divergence of opinions is privacy.
This was especially apparent in the questions about identifying developers.
Interestingly, there was no appreciable difference between general identification of developers who contributed to a project, and specific identification of developers whose work had been analyzed.
The divergent opinions are probably a result of the clash between two ideals: that of openness and attribution, which favors the identification of developers, and that of avoiding harm, which may favor protecting their identity.
Further support for this conjecture is given by the fact that researchers, who are probably less motivated by the ideal of openness, tended to oppose the identification of developers more than the developers themselves.
A workable solution is not to follow a predefined guideline but to ask those you want to identify for their explicit preference and consent.

While our results indicate that researchers generally see things eye to eye with developers, the HC study incident shows that this is not always the case.
It is therefore advisable for open source repositories and projects to draft and publish explicit rules of conduct for researchers who wish to perform research on them.
However, one must remember that regulations only work if there is good will.
For example, Sieber wrote about informed consent that
``a signed consent form is a bureaucratic and legal maneuver that better protects the researcher's institution than it protects the subject'' \cite{sieber01}.
So the real goal is to facilitate a culture of cooperation, not just to draft regulations.

\begin{table}\centering
    \caption{\label{tab:rec}\sl
    Some recommendations for possible ethics guidelines.}
\setlength{\fboxsep}{7pt}
\fbox{\parbox{.92\columnwidth}{\begin{itemize}
    \item Public data, including code and communications, can be used for research.
    \item Projects used in research should be identified; this is based on the tacit assumption that the research does not harm the project.
    \item When excerpts of public data are quoted, their authors should be asked whether they prefer to be identified (for attribution) or not (for privacy).
    \item Developers may be contacted about their work, but the research setting must be disclosed.
    \item Experiments should be cleared with project leaders and must obtain informed consent.
    \item Submitting code to open source projects must only be done in good faith.
    \item Code patches generated by a tool, especially an experimental one, should be noted as such.
    \item License terms should be respected; if this seems to pose a problem, consult with project leaders.
    \item Unsolicited invitations to experiments and surveys should be reduced by
    \begin{itemize}
        \item Targeting only potentially interested developers;
        \item Stopping if the response rate is low (indicating lack of interest);
        \item Stopping when results stabilize (rather than trying to reach a given number of respondents).
    \end{itemize}
    \end{itemize}}}
\end{table}

Some concrete recommendations are given in Table \ref{tab:rec}.
These can be used in three contexts.
The first is research guidelines in open source repositories.
Having such guidelines would ensure better compliance, and avoid the need for creative interpretation of general usage guidelines that do not consider research explicitly (as suggested in \cite{gold22}).
They would also enable large-scale studies on many thousands of projects where it is impractical to verify the preferences of the leaders of each project.
The second is ethics codes by professional societies such as the ACM and IEEE.
Such societies cater not only to practitioners but also to researchers.
They should therefore include research ethics in their guidelines, and they should reach out to affected communities for input about what to include in these guidelines.
Finally, a third context is ethics committees and IRBs charged with approving experiments.
Such committees should consider not only the legal framework, but also the emergent etiquette of the communities from which subjects are recruited.
Communities have opinions and want to be heard.
Our survey is an example of how such relevant guidelines were obtained for the case of the open source community.

A recurring problem is recruiting projects and developers to participate in research \cite{cho99,baltes16}.
Ideally this should be based on an opt-in mechanism to avoid spamming, but such a mechanism does not exist.
The suggestion by Wagner et al.\ that over 30,000 invitations should be sent to obtain 400 respondents for a survey seems excessive;
with such a low response rate they probably have a strong selection bias which invalidates the statistical assumptions.
And such a large number of invitations implies that they are sent over some period of time.
This can be used to reduce the spamming in two ways.
First, if the response rate is found to be very low, this implies that a wide gap exists between what the researchers are interested in and think is important and what invitees care about.
Asking them about things they don't care about makes it spam.
So if the response rate is too low the survey should be stopped.
Second, researchers should analyze the results as they are collected, and discontinue data collection as soon as they appear to stabilize enough for their needs.
In addition, snowballing can be encouraged: if participants agree to forward the invitation to their contacts, this reflects an expectation that they will be interested.
Another idea is that informing invitees of how their email was obtained would be courteous.
Finally, one respondent suggested that invitations should be posted to development mailing list rather than approaching the developers directly, which would be more in line with open source development culture.

\hide{
1) using available data; privacy, confidentiality
2) sending surveys = asking for data in peoples heads; inconvenience
3) performing experiments = asking for help to generate new data; effort
use snowballing to reduce spam and as evidence for interest
need to convince of scientific value
}

\section{Threats to Validity}
\label{sect:threats}

Our survey, like any opinion survey, suffers from a potential threat to construct validity.
Respondents spend only a few seconds forming opinions on hypothetical scenarios that they have not experienced.
For example, developers may not fully realize the risks they take when research is performed on their code, e.g.\ if they are identified and presented in a negative light.
This should be kept in mind especially with regard to privacy and confidentiality.
In addition, as some respondents noted in their comments, survey questions cannot really fully describe a situation, and therefore in many cases the actual answer is ``it depends''.
We note, however, that the vast majority of respondents did make selections from the given options and did not skip questions or select ``other''.

A bigger problem is the threat to external validity, namely whether our results are representative of developers and researchers in general.
This has two facets: whether the sample is big enough, and who is included in it.
Regarding the size of the sample we checked the results obtained from only half our respondents, 88 developers and 22 researchers, and found that the results for developers are essentially the same, and for researchers very close, despite their low number.
It therefore seems that the sample size is not a problem, and the results are representative for developers and researchers who respond to such surveys.

However, external validity may still be compromised due to a possible selection bias.
Our participants reflect a self-selection to accept the invitation and answer the survey.
It is reasonable to assume that practitioners who knew about the Linux-UMN incident --- and especially those who have strong opinions about being experimented on --- had a higher tendency to participate.
Indeed, some of the respondents included rather emotional comments such as ``We are NOT computers'' and ``Consent. It's a thing now. Get it.'' and even much stronger language.
At the same time, those who couldn't care less about ethics most probably just deleted the invitation email and did not participate.
The results may therefore not be representative of the whole population of developers and researchers.
Note, however, that this implies that our results about acceptable behavior may be conservative rather than being too lenient.

\hide{
Also, one participant noted that there was a problem in the DMARC email headers, indicating that the invitation email was not authenticated.
For a highly technical target audience this could cause the invitation to be disregarded, again leading to a bias (the more aware of this, the higher the disregard).
}

\section{Conclusions}
\label{sect:conc}

Both open source developers and software engineering researchers come from a technical background.
As such they may have blind spots when it comes to social issues and to ethics.
As Harrison wrote,
``Physicists don’t have to ask an electron if they can measure it, nor are they obligated to allow the electron to quit the experiment at any time'' \cite{harrison00}.
And awareness of this often leads to a reliance on (and confinement to) legal requirements and licenses \cite{menlo,gold22}.
But the laws originate from a background of privacy issues, and the licenses from a background of code distribution, so their implications regarding ethics issues like consent are incidental rather than intended.

Current ethics guidelines were defined in the context of bio-medical research, and their application to software engineering research requires some adjustments.
For example, issues that are not well covered include
\begin{itemize}
    \item Using existing publicly accessible artifacts (code, development history, documentation, and communications among developers)
    \item Observational studies (watching people work)
    \item Interacting with developers workflows (contributing code, contributing tools, collaborating in other ways)
\end{itemize}
One approach to reduce ethical friction is therefore to develop guidelines that are better aligned with the practices of software development.
Some ideas along this line were proposed above in Table \ref{tab:rec}.

Another possible approach is to join forces:
instead of imposing on the research subjects and potentially alienating them, involve them as participants in the research \cite{bakardjieva00}.
Such an approach is in line with the open source philosophy, and may be expected to lead to better scientific results --- results that are more correct and more relevant, being based on the developers' point of view.
A further step is this direction is to use participatory research:
given that many researchers are also contributing developers, they can study the projects they work on from within.
This proactively returns to the community \cite{oezbek08}, in a way that may be better appreciated than the publication of an academic paper.

Yet another alternative is to try to use industrial collaboration \cite{rico21}.
Such collaborations naturally enjoy high relevance, because they necessarily focus on real needs.
In addition, for smaller research projects one can hire developers for experiments \cite{sjoberg02,sjoberg03}, or even use ``human computation'' platforms like Mechanical Turk \cite{sabou20}.
Like experiments on open source projects, these approaches have the advantage of having developers work in their normal environment.

Last, the reaction to ethics violations should be carefully considered.
Research like the HC study is important, and should not be discounted outright.
In this specific case, its main contribution was to show how potential vulnerabilities could be turned into real vulnerabilities, and suggest that this could be hidden in innocent-looking patches.
Another unintended contribution was to show that the Linux vetting procedure works, and in fact prevented these innocent-looking commits from being accepted.
But the study suffered from a significant ethics blind spot.
To be acceptable, it should have been coordinated with the target project, and performed in a manner they approve.
To prevent such incidents from repeating, we do not need to chastise UMN --- we need to develop procedures and mechanisms to coordinate research on open-source projects.

\hide{
Developers, especially in free and open-source, deserve ethical treatment.
But at the same time, we should all remember to also give ethical treatment to users and the public in general.
It seems that too often developers are blind to their violations of basic norms and values, and to the possible harm they inflict \cite{obie21}.
The appeal for ethical treatment in research will be stronger if applications are ethical themselves.
}

\subsection*{Data Availability}

The full responses to the survey are available on Zenodo using DOI
10.5281/zenodo.5752053.

\subsection*{Acknowledgments}

Many thanks are due to all the survey participants, especially those who invested extra effort to write comments and explain their positions.

\bibliographystyle{myabbrv}
\bibliography{abbrv,se,misc,par}


\end{document}

%% file: plt-43.tex
\vspace{1mm}
\noindent\begin{tabular}{@{}p{86mm}}
\hline
\rule{0pt}{2ex}Q43\footnotemark: Status (check all that apply):\\

\multicolumn{1}{@{}r@{}}{

\adjustbox{valign=t}{\begin{tikzpicture}[scale=0.5]
 \begin{axis}[
   font=\sffamily\Large,
   width=120pt,
   height=220pt,
   scale only axis=true,
   xbar,
   bar width=8pt,
   ytick=data,
   yticklabel style={font=\sffamily\LARGE},
  yticklabels={Other,Student,Researcher,Management position,Program as a hobby (unpaid),Professional developer (paid employee),Freelance developer (paid per project)},
   enlarge y limits=0.1,
   xlabel = {Percents},
   xmin=0,xmax=110,
   xtick={0,20,40,60,80,100},
  ]
   \addplot[fill=blue,color=blue!60,mark=none] coordinates {(5.98802395209581,1) (11.9760479041916,2) (16.1676646706587,3) (16.7664670658683,4) (52.0958083832335,5) (75.4491017964072,6) (22.7544910179641,7) };
  \addplot[fill=orange,color=orange!60,mark=none] coordinates {(1.81818181818182,1) (7.27272727272727,2) (100,3) (7.27272727272727,4) (18.1818181818182,5) (5.45454545454545,6) (0,7) };
  \legend{Dev.,Res.}
 \end{axis}
\end{tikzpicture}}
}\\
\hline

\end{tabular}
\vspace{2mm}
\footnotetext{The question numbers indicate the order in which they were presented to participants, which is different from the order used here.
In particular, the demographic questions originally appeared at the end.}

%% file: plt-39.tex
\vspace{1mm}
\noindent\begin{tabular}{@{}p{45mm}p{37mm}}
\hline

\adjustbox{valign=t}{\begin{tikzpicture}[scale=0.5]
 \begin{axis}[
   font=\sffamily\Large,
   width=180pt,
   height=120pt,
   scale only axis=true,
   ybar,
   bar width=7.5pt,
   xtick=data,
   ylabel = {Cumulative percent},
   ymin=0,
  ymax=100,
  ytick={0,20,40,60,80,100},
  xmin=0,
  xmax=50,
  xtick={0,10,20,30,40,50},
  title=Q39: Dev. experience years,
  title style={font=\rmfamily\LARGE},
legend pos=south east,
  ]
  \addplot[sharp plot,blue,line width=1.3] coordinates {(0,1.79640718562874) (1,3.59281437125748) (2,8.38323353293413) (4,12.5748502994012) (5,20.9580838323353) (6,23.3532934131736) (7,27.5449101796407) (8,32.9341317365269) (9,34.1317365269461) (10,47.3053892215569) (11,49.7005988023952) (12,52.6946107784431) (13,56.8862275449102) (14,57.4850299401198) (15,65.2694610778443) (16,67.065868263473) (17,69.4610778443114) (18,71.2574850299401) (20,80.2395209580838) (21,80.8383233532934) (22,81.437125748503) (23,82.6347305389221) (25,87.4251497005988) (26,88.0239520958084) (28,88.6227544910179) (29,89.2215568862275) (30,96.4071856287425) (35,97.0059880239521) (40,98.2035928143712) (41,98.8023952095808) (46,99.4011976047904) (50,99.9999999999999) };
  \addplot[sharp plot,orange,line width=1.3] coordinates {(0,5.76923076923077) (1,13.4615384615385) (2,19.2307692307692) (3,23.0769230769231) (5,38.4615384615385) (6,40.3846153846154) (8,44.2307692307692) (9,46.1538461538461) (10,57.6923076923077) (12,59.6153846153846) (15,73.0769230769231) (16,75) (20,90.3846153846154) (25,92.3076923076923) (30,94.2307692307692) (35,96.1538461538461) (36,98.0769230769231) (44,100) };
  \legend{Dev.,Res.}
 \end{axis}
\end{tikzpicture}}
 & 
\adjustbox{valign=t}{\begin{tikzpicture}[scale=0.5]
 \begin{axis}[
   font=\sffamily\Large,
   width=180pt,
   height=120pt,
   scale only axis=true,
   ybar,
   bar width=7.5pt,
   xtick=data,
   ylabel = {Cumulative percent},
   ymin=0,
  ymax=100,
  ytick={0,20,40,60,80,100},
  xmin=0,
  xmax=100,
  xtick={0,20,40,60,80,100},
  title=Q42: Papers published,
  title style={font=\rmfamily\LARGE},
legend pos=south east,
  ]
  \addplot[sharp plot,blue,line width=1.3] coordinates {(0,77.639751552795) (1,84.472049689441) (2,88.1987577639752) (3,92.5465838509317) (4,95.0310559006211) (5,97.5155279503106) (8,98.1366459627329) (10,100) };
  \addplot[sharp plot,orange,line width=1.3] coordinates {(0,13.4615384615385) (1,17.3076923076923) (2,19.2307692307692) (4,21.1538461538462) (5,28.8461538461538) (8,30.7692307692308) (10,40.3846153846154) (15,46.1538461538462) (20,55.7692307692308) (25,59.6153846153846) (30,67.3076923076923) (40,73.0769230769231) (50,82.6923076923077) (70,88.4615384615385) (80,90.3846153846154) (90,92.3076923076923) (100,98.0769230769231) };
  \legend{Dev.,Res.}
 \end{axis}
\end{tikzpicture}}
\\
\hline

\end{tabular}
\vspace{2mm}

%% file: plt-40.tex
\vspace{1mm}
\noindent\begin{tabular}{@{}p{45mm}p{37mm}}
\hline

\adjustbox{valign=t}{\begin{tikzpicture}[scale=0.5]
 \begin{axis}[
   font=\sffamily\Large,
   width=180pt,
   height=120pt,
   scale only axis=true,
   ybar,
   bar width=7.5pt,
   xtick=data,
   ylabel = {Cumulative percent},
   ymin=0,
  ymax=100,
  ytick={0,20,40,60,80,100},
  xmin=0,
  xmax=40,
  xtick={0,10,20,30,40},
  title=Q40: Open source years,
  title style={font=\rmfamily\LARGE},
legend pos=south east,
  ]
  \addplot[sharp plot,blue,line width=1.3] coordinates {(0,0.602409638554217) (1,3.6144578313253) (2,9.63855421686747) (3,13.855421686747) (4,20.4819277108434) (5,31.3253012048193) (6,38.5542168674699) (7,48.7951807228916) (8,53.6144578313253) (9,54.2168674698795) (10,66.2650602409639) (11,67.4698795180723) (12,68.6746987951807) (13,71.6867469879518) (14,72.8915662650602) (15,82.5301204819277) (16,83.1325301204819) (17,84.3373493975904) (18,85.5421686746988) (20,94.578313253012) (23,95.7831325301205) (24,96.3855421686747) (25,98.7951807228916) (30,99.3975903614458) (40,100) };
  \addplot[sharp plot,orange,line width=1.3] coordinates {(0,37.037037037037) (1,44.4444444444444) (2,48.1481481481481) (3,55.5555555555556) (5,62.962962962963) (6,66.6666666666667) (7,70.3703703703704) (8,77.7777777777778) (9,79.6296296296296) (10,88.8888888888889) (14,90.7407407407407) (15,94.4444444444444) (20,98.1481481481482) (22,100) };
  \legend{Dev.,Res.}
 \end{axis}
\end{tikzpicture}}
 & 
\adjustbox{valign=t}{\begin{tikzpicture}[scale=0.5]
 \begin{axis}[
   font=\sffamily\Large,
   width=180pt,
   height=120pt,
   scale only axis=true,
   ybar,
   bar width=7.5pt,
   xtick=data,
   ylabel = {Cumulative percent},
   ymin=0,
  ymax=100,
  ytick={0,20,40,60,80,100},
  xmin=0,
  xmax=100,
  xtick={0,20,40,60,80,100},
  title=Q41: Open source projects,
  title style={font=\rmfamily\LARGE},
legend pos=south east,
  ]
  \addplot[sharp plot,blue,line width=1.3] coordinates {(0,0.602409638554217) (1,3.01204819277108) (2,10.2409638554217) (3,23.4939759036145) (4,29.5180722891566) (5,42.7710843373494) (6,46.9879518072289) (7,47.5903614457831) (8,48.1927710843373) (10,69.2771084337349) (12,71.6867469879518) (14,72.289156626506) (15,77.710843373494) (20,86.7469879518072) (22,87.3493975903615) (25,87.9518072289157) (30,90.9638554216868) (36,91.566265060241) (40,92.1686746987952) (50,95.7831325301205) (100,97.5903614457832) };
  \addplot[sharp plot,orange,line width=1.3] coordinates {(0,30.7692307692308) (1,40.3846153846154) (2,46.1538461538462) (3,51.9230769230769) (4,55.7692307692308) (5,71.1538461538461) (6,76.9230769230769) (10,88.4615384615385) (15,90.3846153846154) (17,92.3076923076923) (20,98.0769230769231) (50,100) };
  \legend{Dev.,Res.}
 \end{axis}
\end{tikzpicture}}
\\
\hline

\end{tabular}
\vspace{2mm}

%% file: plt-1.tex
\vspace{1mm}
\noindent\begin{tabular}{@{}p{86mm}}
\hline
\rule{0pt}{2ex}Q1: Did you hear of the Linux-UMN incident?\\

\multicolumn{1}{@{}r@{}}{

\adjustbox{valign=t}{\begin{tikzpicture}[scale=0.5]
 \begin{axis}[
   font=\sffamily\Large,
   width=120pt,
   height=100pt,
   scale only axis=true,
   xbar,
   bar width=8pt,
   ytick=data,
   yticklabel style={font=\sffamily\LARGE},
  yticklabels={Did not hear of it,Heard but did not know details,Followed it when it happened},
   enlarge y limits=0.3,
   xlabel = {Percents},
   xmin=0,xmax=60,
   xtick={0,10,20,30,40,50,60},
   legend pos=south east
  ]
   \addplot[fill=blue,color=blue!60,mark=none] coordinates {(27.3809523809524,1) (13.6904761904762,2) (58.9285714285714,3) };
  \addplot[fill=orange,color=orange!60,mark=none] coordinates {(26.7857142857143,1) (21.4285714285714,2) (51.7857142857143,3) };
  \legend{Dev.,Res.}
 \end{axis}
\end{tikzpicture}}
}\\
\hline

\end{tabular}
\vspace{2mm}

%% file: plt-2.tex
\vspace{1mm}
\noindent\begin{tabular}{@{}p{86mm}}
\hline
\rule{0pt}{2ex}Q2: What do you think of the decision to ban UMN from contributing to the Linux kernel?\\

\multicolumn{1}{@{}r@{}}{

\adjustbox{valign=t}{\begin{tikzpicture}[scale=0.5]
 \begin{axis}[
   font=\sffamily\Large,
   width=120pt,
   height=100pt,
   scale only axis=true,
   xbar,
   bar width=8pt,
   ytick=data,
   yticklabel style={font=\sffamily\LARGE},
  yticklabels={Not justified and blown out of proportion,Somewhat exaggerated; more nuanced better,It was justified: such behavior is a big deal},
   enlarge y limits=0.3,
   xlabel = {Percents},
   xmin=0,xmax=70,
   xtick={0,20,40,60},
   legend pos=south east
  ]
   \addplot[fill=blue,color=blue!60,mark=none] coordinates {(2.38095238095238,1) (32.7380952380952,2) (64.8809523809524,3) };
  \addplot[fill=orange,color=orange!60,mark=none] coordinates {(1.81818181818182,1) (40,2) (58.1818181818182,3) };
  \legend{Dev.,Res.}
 \end{axis}
\end{tikzpicture}}
}\\
\hline

\end{tabular}
\vspace{2mm}

%% file: plt-3.tex
\vspace{1mm}
\noindent\begin{tabular}{@{}p{86mm}}
\hline
\rule{0pt}{2ex}Q3: The UMN IRB (Institutional Review Board) had given an exemption to this research based on the perception that studying the kernel patch process is not human research. Do you agree with this judgement?\\

\multicolumn{1}{@{}r@{}}{

\adjustbox{valign=t}{\begin{tikzpicture}[scale=0.5]
 \begin{axis}[
   font=\sffamily\Large,
   width=120pt,
   height=100pt,
   scale only axis=true,
   xbar,
   bar width=8pt,
   ytick=data,
   yticklabel style={font=\sffamily\LARGE},
  yticklabels={Indeed not human research,Hard to tell,Definitely human research},
   enlarge y limits=0.3,
   xlabel = {Percents},
   xmin=0,xmax=60,
   xtick={0,10,20,30,40,50,60},
   legend pos=south east
  ]
   \addplot[fill=blue,color=blue!60,mark=none] coordinates {(13.9393939393939,1) (43.030303030303,2) (43.030303030303,3) };
  \addplot[fill=orange,color=orange!60,mark=none] coordinates {(12.5,1) (33.9285714285714,2) (53.5714285714286,3) };
  \legend{Dev.,Res.}
 \end{axis}
\end{tikzpicture}}
}\\
\hline

\end{tabular}
\vspace{2mm}

%% file: plt-4.tex
\vspace{1mm}
\noindent\begin{tabular}{@{}p{86mm}}
\hline
\rule{0pt}{2ex}Q4: What was the worst offense in the UMN study?\\

\multicolumn{1}{@{}r@{}}{

\adjustbox{valign=t}{\begin{tikzpicture}[scale=0.5]
 \begin{axis}[
   font=\sffamily\Large,
   width=120pt,
   height=160pt,
   scale only axis=true,
   xbar,
   bar width=8pt,
   ytick=data,
   yticklabel style={font=\sffamily\LARGE},
  yticklabels={Other,The study was fine: there was no offence,Risk of buggy code entering the distribution,Wasting the time of Linux maintainers,Treating Linux developers as guinea pigs},
   enlarge y limits=0.2,
   xlabel = {Percents},
   xmin=0,xmax=50,
   xtick={0,10,20,30,40,50},
   legend pos=south east
  ]
   \addplot[fill=blue,color=blue!60,mark=none] coordinates {(1.21212121212121,1) (1.81818181818182,2) (43.8383838383838,3) (31.7171717171717,4) (21.4141414141414,5) };
  \addplot[fill=orange,color=orange!60,mark=none] coordinates {(1.78571428571429,1) (0,2) (32.7380952380952,3) (41.6666666666667,4) (23.8095238095238,5) };
  \legend{Dev.,Res.}
 \end{axis}
\end{tikzpicture}}
}\\
\hline

\end{tabular}
\vspace{2mm}

%% file: plt-5.tex
\vspace{1mm}
\noindent\begin{tabular}{@{}p{86mm}}
\hline
\rule{0pt}{2ex}Q5: With regard to Kroah-Hartman's comment that "Our community does not appreciate being experimented on", is it at all possible to conduct an ethical experiment on whether open-source maintainers detect buggy code contributions?\\

\multicolumn{1}{@{}r@{}}{

\adjustbox{valign=t}{\begin{tikzpicture}[scale=0.5]
 \begin{axis}[
   font=\sffamily\Large,
   width=120pt,
   height=220pt,
   scale only axis=true,
   xbar,
   bar width=8pt,
   ytick=data,
   yticklabel style={font=\sffamily\LARGE},
  yticklabels={Other,Experiment was OK; buggy code not distributed,OK if cleared in advance with project leaders,OK with dev's who identify as willing to participate,OK with informed consent; risks harming validity,Experiment on buggy contributions not ethical,All experiments on developers are unethical},
   enlarge y limits=0.1,
   xlabel = {Percents},
   xmin=0,xmax=50,
   xtick={0,10,20,30,40,50},
   legend pos=south east
  ]
   \addplot[fill=blue,color=blue!60,mark=none] coordinates {(4.84848484848485,1) (6.66666666666667,2) (39.3939393939394,3) (15.1515151515152,4) (17.5757575757576,5) (12.7272727272727,6) (4.24242424242424,7) };
  \addplot[fill=orange,color=orange!60,mark=none] coordinates {(7.27272727272727,1) (0,2) (21.8181818181818,3) (23.6363636363636,4) (41.8181818181818,5) (1.81818181818182,6) (3.63636363636364,7) };
  \legend{Dev.,Res.}
 \end{axis}
\end{tikzpicture}}
}\\
\hline

\end{tabular}
\vspace{2mm}

%% file: plt-6.tex
\vspace{1mm}
\noindent\begin{tabular}{@{}p{45mm}p{37mm}}
\hline
\rule{0pt}{2ex}\raggedright Q6: Using code examples from the project without asking permission&

\adjustbox{valign=t}{\begin{tikzpicture}[scale=0.5]
 \begin{axis}[
   font=\sffamily\Large,
   width=180pt,
   height=120pt,
   scale only axis=true,
   ybar,
   bar width=7.5pt,
   xtick=data,
   ylabel = {Percents},
   ymin=0,ymax=100,
   ytick={0,20,40,60,80,100},
   legend style={at={(0.98,0.5)},anchor=east}
  ]
  \addplot[fill=blue,color=blue!60,mark=none] coordinates {(1,52.6946107784431) (2,16.7664670658683) (3,10.1796407185629) (4,7.78443113772455) (5,7.18562874251497) (6,2.9940119760479) (7,2.39520958083832) };
  \addplot[fill=orange,color=orange!60,mark=none] coordinates {(1,62.962962962963) (2,12.962962962963) (3,0) (4,11.1111111111111) (5,7.40740740740741) (6,1.85185185185185) (7,3.7037037037037) };
	\addplot[sharp plot,blue,line width=1.3] coordinates {(1,52.6946107784431) (2,69.4610778443114) (3,79.6407185628743) (4,87.4251497005988) (5,94.6107784431138) (6,97.6047904191617) (7,100) };
	\addplot[sharp plot,orange,line width=1.3] coordinates {(1,62.962962962963) (2,75.9259259259259) (3,75.9259259259259) (4,87.037037037037) (5,94.4444444444444) (6,96.2962962962963) (7,100) };
  \legend{Dev.,Res.}
 \end{axis}
\end{tikzpicture}}
\\
\hline
\rule{0pt}{2ex}\raggedright Q7: Publishing open-source code examples from the project in a copyrighted article&

\adjustbox{valign=t}{\begin{tikzpicture}[scale=0.5]
 \begin{axis}[
   font=\sffamily\Large,
   width=180pt,
   height=120pt,
   scale only axis=true,
   ybar,
   bar width=7.5pt,
   xtick=data,
   ylabel = {Percents},
   ymin=0,ymax=100,
   ytick={0,20,40,60,80,100},
   legend style={at={(0.98,0.4)},anchor=east}
  ]
  \addplot[fill=blue,color=blue!60,mark=none] coordinates {(1,30.3030303030303) (2,12.1212121212121) (3,9.09090909090909) (4,9.6969696969697) (5,14.5454545454545) (6,12.1212121212121) (7,12.1212121212121) };
  \addplot[fill=orange,color=orange!60,mark=none] coordinates {(1,48.1481481481481) (2,11.1111111111111) (3,9.25925925925926) (4,9.25925925925926) (5,11.1111111111111) (6,11.1111111111111) (7,0) };
	\addplot[sharp plot,blue,line width=1.3] coordinates {(1,30.3030303030303) (2,42.4242424242424) (3,51.5151515151515) (4,61.2121212121212) (5,75.7575757575758) (6,87.8787878787879) (7,100) };
	\addplot[sharp plot,orange,line width=1.3] coordinates {(1,48.1481481481481) (2,59.2592592592593) (3,68.5185185185185) (4,77.7777777777778) (5,88.8888888888889) (6,100) (7,100) };
  \legend{Dev.,Res.}
 \end{axis}
\end{tikzpicture}}
\\
\hline

\end{tabular}
\vspace{2mm}

%% file: plt-8.tex
\vspace{1mm}
\noindent\begin{tabular}{@{}p{45mm}p{37mm}}
\hline
\rule{0pt}{2ex}\raggedright Q8: Reading and analyzing the text of communications between developers of the project to better understand technical issues&

\adjustbox{valign=t}{\begin{tikzpicture}[scale=0.5]
 \begin{axis}[
   font=\sffamily\Large,
   width=180pt,
   height=120pt,
   scale only axis=true,
   ybar,
   bar width=7.5pt,
   xtick=data,
   ylabel = {Percents},
   ymin=0,ymax=100,
   ytick={0,20,40,60,80,100},
   legend style={at={(0.98,0.5)},anchor=east}
  ]
  \addplot[fill=blue,color=blue!60,mark=none] coordinates {(1,63.2530120481928) (2,20.4819277108434) (3,7.2289156626506) (4,6.02409638554217) (5,1.20481927710843) (6,1.20481927710843) (7,0.602409638554217) };
  \addplot[fill=orange,color=orange!60,mark=none] coordinates {(1,58.1818181818182) (2,9.09090909090909) (3,10.9090909090909) (4,12.7272727272727) (5,5.45454545454545) (6,1.81818181818182) (7,1.81818181818182) };
	\addplot[sharp plot,blue,line width=1.3] coordinates {(1,63.2530120481928) (2,83.7349397590361) (3,90.9638554216867) (4,96.9879518072289) (5,98.1927710843373) (6,99.3975903614458) (7,100) };
	\addplot[sharp plot,orange,line width=1.3] coordinates {(1,58.1818181818182) (2,67.2727272727273) (3,78.1818181818182) (4,90.9090909090909) (5,96.3636363636364) (6,98.1818181818182) (7,100) };
  \legend{Dev.,Res.}
 \end{axis}
\end{tikzpicture}}
\\
\hline
\rule{0pt}{2ex}\raggedright Q9: Reading and analyzing the text of communications between developers of the project to better understand the social interactions between them&

\adjustbox{valign=t}{\begin{tikzpicture}[scale=0.5]
 \begin{axis}[
   font=\sffamily\Large,
   width=180pt,
   height=120pt,
   scale only axis=true,
   ybar,
   bar width=7.5pt,
   xtick=data,
   ylabel = {Percents},
   ymin=0,ymax=100,
   ytick={0,20,40,60,80,100},
   legend style={at={(0.98,0.5)},anchor=east}
  ]
  \addplot[fill=blue,color=blue!60,mark=none] coordinates {(1,45.5089820359281) (2,17.3652694610778) (3,14.3712574850299) (4,8.38323353293413) (5,7.18562874251497) (6,4.79041916167665) (7,2.39520958083832) };
  \addplot[fill=orange,color=orange!60,mark=none] coordinates {(1,47.2727272727273) (2,10.9090909090909) (3,12.7272727272727) (4,16.3636363636364) (5,5.45454545454545) (6,5.45454545454545) (7,1.81818181818182) };
	\addplot[sharp plot,blue,line width=1.3] coordinates {(1,45.5089820359281) (2,62.874251497006) (3,77.2455089820359) (4,85.6287425149701) (5,92.814371257485) (6,97.6047904191617) (7,100) };
	\addplot[sharp plot,orange,line width=1.3] coordinates {(1,47.2727272727273) (2,58.1818181818182) (3,70.9090909090909) (4,87.2727272727273) (5,92.7272727272727) (6,98.1818181818182) (7,100) };
  \legend{Dev.,Res.}
 \end{axis}
\end{tikzpicture}}
\\
\hline

\end{tabular}
\vspace{2mm}

%% file: plt-10.tex
\vspace{1mm}
\noindent\begin{tabular}{@{}p{45mm}p{37mm}}
\hline
\rule{0pt}{2ex}\raggedright Q10: Using machine learning to analyze the text of all communications between developers of the project and derive statistical characterizations (e.g. "73\% of comments were negative")&

\adjustbox{valign=t}{\begin{tikzpicture}[scale=0.5]
 \begin{axis}[
   font=\sffamily\Large,
   width=180pt,
   height=120pt,
   scale only axis=true,
   ybar,
   bar width=7.5pt,
   xtick=data,
   ylabel = {Percents},
   ymin=0,ymax=100,
   ytick={0,20,40,60,80,100},
   legend style={at={(0.98,0.5)},anchor=east}
  ]
  \addplot[fill=blue,color=blue!60,mark=none] coordinates {(1,37.9518072289157) (2,19.8795180722892) (3,13.855421686747) (4,9.03614457831325) (5,9.63855421686747) (6,4.81927710843374) (7,4.81927710843374) };
  \addplot[fill=orange,color=orange!60,mark=none] coordinates {(1,47.2727272727273) (2,16.3636363636364) (3,14.5454545454545) (4,12.7272727272727) (5,3.63636363636364) (6,3.63636363636364) (7,1.81818181818182) };
	\addplot[sharp plot,blue,line width=1.3] coordinates {(1,37.9518072289157) (2,57.8313253012048) (3,71.6867469879518) (4,80.7228915662651) (5,90.3614457831325) (6,95.1807228915663) (7,100) };
	\addplot[sharp plot,orange,line width=1.3] coordinates {(1,47.2727272727273) (2,63.6363636363636) (3,78.1818181818182) (4,90.9090909090909) (5,94.5454545454545) (6,98.1818181818182) (7,100) };
  \legend{Dev.,Res.}
 \end{axis}
\end{tikzpicture}}
\\
\hline

\end{tabular}
\vspace{2mm}

%% file: plt-18.tex
\vspace{1mm}
\noindent\begin{tabular}{@{}p{45mm}p{37mm}}
\hline
\rule{0pt}{2ex}\raggedright Q18: Identifying developers who contributed to the project in a paper about the research&

\adjustbox{valign=t}{\begin{tikzpicture}[scale=0.5]
 \begin{axis}[
   font=\sffamily\Large,
   width=180pt,
   height=120pt,
   scale only axis=true,
   ybar,
   bar width=7.5pt,
   xtick=data,
   ylabel = {Percents},
   ymin=0,ymax=100,
   ytick={0,20,40,60,80,100},
   legend pos=north west
  ]
  \addplot[fill=blue,color=blue!60,mark=none] coordinates {(1,20.2453987730061) (2,15.9509202453988) (3,16.5644171779141) (4,14.7239263803681) (5,9.8159509202454) (6,6.13496932515337) (7,16.5644171779141) };
  \addplot[fill=orange,color=orange!60,mark=none] coordinates {(1,12.7272727272727) (2,9.09090909090909) (3,7.27272727272727) (4,9.09090909090909) (5,10.9090909090909) (6,12.7272727272727) (7,38.1818181818182) };
	\addplot[sharp plot,blue,line width=1.3] coordinates {(1,20.2453987730061) (2,36.1963190184049) (3,52.760736196319) (4,67.4846625766871) (5,77.3006134969325) (6,83.4355828220859) (7,100) };
	\addplot[sharp plot,orange,line width=1.3] coordinates {(1,12.7272727272727) (2,21.8181818181818) (3,29.0909090909091) (4,38.1818181818182) (5,49.0909090909091) (6,61.8181818181818) (7,100) };
  \legend{Dev.,Res.}
 \end{axis}
\end{tikzpicture}}
\\
\hline
\rule{0pt}{2ex}\raggedright Q19: Identifying developers who wrote specific code or comments that were quoted in a paper about the research&

\adjustbox{valign=t}{\begin{tikzpicture}[scale=0.5]
 \begin{axis}[
   font=\sffamily\Large,
   width=180pt,
   height=120pt,
   scale only axis=true,
   ybar,
   bar width=7.5pt,
   xtick=data,
   ylabel = {Percents},
   ymin=0,ymax=100,
   ytick={0,20,40,60,80,100},
   legend pos=north west
  ]
  \addplot[fill=blue,color=blue!60,mark=none] coordinates {(1,17.5757575757576) (2,13.9393939393939) (3,15.1515151515152) (4,12.7272727272727) (5,13.3333333333333) (6,6.06060606060606) (7,21.2121212121212) };
  \addplot[fill=orange,color=orange!60,mark=none] coordinates {(1,12.7272727272727) (2,9.09090909090909) (3,9.09090909090909) (4,7.27272727272727) (5,10.9090909090909) (6,16.3636363636364) (7,34.5454545454545) };
	\addplot[sharp plot,blue,line width=1.3] coordinates {(1,17.5757575757576) (2,31.5151515151515) (3,46.6666666666667) (4,59.3939393939394) (5,72.7272727272727) (6,78.7878787878788) (7,100) };
	\addplot[sharp plot,orange,line width=1.3] coordinates {(1,12.7272727272727) (2,21.8181818181818) (3,30.9090909090909) (4,38.1818181818182) (5,49.0909090909091) (6,65.4545454545455) (7,100) };
  \legend{Dev.,Res.}
 \end{axis}
\end{tikzpicture}}
\\
\hline
\rule{0pt}{2ex}\raggedright Q21: Voicing an opinion about the quality of the work of specific identified developers&

\adjustbox{valign=t}{\begin{tikzpicture}[scale=0.5]
 \begin{axis}[
   font=\sffamily\Large,
   width=180pt,
   height=120pt,
   scale only axis=true,
   ybar,
   bar width=7.5pt,
   xtick=data,
   ylabel = {Percents},
   ymin=0,ymax=100,
   ytick={0,20,40,60,80,100},
   legend pos=north west
  ]
  \addplot[fill=blue,color=blue!60,mark=none] coordinates {(1,19.6319018404908) (2,9.20245398773006) (3,12.8834355828221) (4,9.8159509202454) (5,14.7239263803681) (6,14.1104294478528) (7,19.6319018404908) };
  \addplot[fill=orange,color=orange!60,mark=none] coordinates {(1,10.9090909090909) (2,1.81818181818182) (3,5.45454545454545) (4,12.7272727272727) (5,9.09090909090909) (6,18.1818181818182) (7,41.8181818181818) };
	\addplot[sharp plot,blue,line width=1.3] coordinates {(1,19.6319018404908) (2,28.8343558282209) (3,41.7177914110429) (4,51.5337423312883) (5,66.2576687116564) (6,80.3680981595092) (7,100) };
	\addplot[sharp plot,orange,line width=1.3] coordinates {(1,10.9090909090909) (2,12.7272727272727) (3,18.1818181818182) (4,30.9090909090909) (5,40) (6,58.1818181818182) (7,100) };
  \legend{Dev.,Res.}
 \end{axis}
\end{tikzpicture}}
\\
\hline
\rule{0pt}{2ex}\raggedright Q20: Voicing an opinion about the quality of the work on the project&

\adjustbox{valign=t}{\begin{tikzpicture}[scale=0.5]
 \begin{axis}[
   font=\sffamily\Large,
   width=180pt,
   height=120pt,
   scale only axis=true,
   ybar,
   bar width=7.5pt,
   xtick=data,
   ylabel = {Percents},
   ymin=0,ymax=100,
   ytick={0,20,40,60,80,100},
   legend style={at={(0.98,0.45)},anchor=east}
  ]
  \addplot[fill=blue,color=blue!60,mark=none] coordinates {(1,39.7590361445783) (2,19.2771084337349) (3,12.6506024096386) (4,11.4457831325301) (5,7.2289156626506) (6,7.2289156626506) (7,2.40963855421687) };
  \addplot[fill=orange,color=orange!60,mark=none] coordinates {(1,23.6363636363636) (2,25.4545454545455) (3,10.9090909090909) (4,12.7272727272727) (5,7.27272727272727) (6,14.5454545454545) (7,5.45454545454545) };
	\addplot[sharp plot,blue,line width=1.3] coordinates {(1,39.7590361445783) (2,59.0361445783133) (3,71.6867469879518) (4,83.1325301204819) (5,90.3614457831325) (6,97.5903614457831) (7,100) };
	\addplot[sharp plot,orange,line width=1.3] coordinates {(1,23.6363636363636) (2,49.0909090909091) (3,60) (4,72.7272727272727) (5,80) (6,94.5454545454545) (7,100) };
  \legend{Dev.,Res.}
 \end{axis}
\end{tikzpicture}}
\\
\hline

\end{tabular}
\vspace{2mm}

%% file: plt-11.tex
\vspace{1mm}
\noindent\begin{tabular}{@{}p{45mm}p{37mm}}
\hline
\rule{0pt}{2ex}\raggedright Q11: Approaching developers to ask them about their code or the considerations which guided its writing&

\adjustbox{valign=t}{\begin{tikzpicture}[scale=0.5]
 \begin{axis}[
   font=\sffamily\Large,
   width=180pt,
   height=120pt,
   scale only axis=true,
   ybar,
   bar width=7.5pt,
   xtick=data,
   ylabel = {Percents},
   ymin=0,ymax=100,
   ytick={0,20,40,60,80,100},
   legend style={at={(0.98,0.5)},anchor=east}
  ]
  \addplot[fill=blue,color=blue!60,mark=none] coordinates {(1,71.2574850299401) (2,16.1676646706587) (3,4.79041916167665) (4,4.79041916167665) (5,2.9940119760479) (6,0) (7,0) };
  \addplot[fill=orange,color=orange!60,mark=none] coordinates {(1,65.4545454545455) (2,16.3636363636364) (3,7.27272727272727) (4,3.63636363636364) (5,1.81818181818182) (6,5.45454545454545) (7,0) };
	\addplot[sharp plot,blue,line width=1.3] coordinates {(1,71.2574850299401) (2,87.4251497005988) (3,92.2155688622754) (4,97.0059880239521) (5,100) (6,100) (7,100) };
	\addplot[sharp plot,orange,line width=1.3] coordinates {(1,65.4545454545455) (2,81.8181818181818) (3,89.0909090909091) (4,92.7272727272727) (5,94.5454545454545) (6,100) (7,100) };
  \legend{Dev.,Res.}
 \end{axis}
\end{tikzpicture}}
\\
\hline
\end{tabular}

\noindent\begin{tabular}{@{}p{45mm}p{37mm}}
\hline
\rule{0pt}{2ex}\raggedright Q12: Engaging developers without explaining that this is part of an experiment and obtaining informed consent to participate&

\adjustbox{valign=t}{\begin{tikzpicture}[scale=0.5]
 \begin{axis}[
   font=\sffamily\Large,
   width=180pt,
   height=120pt,
   scale only axis=true,
   ybar,
   bar width=7.5pt,
   xtick=data,
   ylabel = {Percents},
   ymin=0,ymax=100,
   ytick={0,20,40,60,80,100},
   legend pos=north west
  ]
  \addplot[fill=blue,color=blue!60,mark=none] coordinates {(1,2.9940119760479) (2,5.38922155688623) (3,7.78443113772455) (4,20.3592814371257) (5,17.3652694610778) (6,19.1616766467066) (7,26.9461077844311) };
  \addplot[fill=orange,color=orange!60,mark=none] coordinates {(1,3.63636363636364) (2,7.27272727272727) (3,0) (4,14.5454545454545) (5,16.3636363636364) (6,21.8181818181818) (7,36.3636363636364) };
	\addplot[sharp plot,blue,line width=1.3] coordinates {(1,2.9940119760479) (2,8.38323353293413) (3,16.1676646706587) (4,36.5269461077844) (5,53.8922155688623) (6,73.0538922155689) (7,100) };
	\addplot[sharp plot,orange,line width=1.3] coordinates {(1,3.63636363636364) (2,10.9090909090909) (3,10.9090909090909) (4,25.4545454545455) (5,41.8181818181818) (6,63.6363636363636) (7,100) };
  \legend{Dev.,Res.}
 \end{axis}
\end{tikzpicture}}
\\
\hline
\rule{0pt}{2ex}\raggedright Q13: Telling developers that this is an experiment, but deceiving them about the details so as not to affect their behavior&

\adjustbox{valign=t}{\begin{tikzpicture}[scale=0.5]
 \begin{axis}[
   font=\sffamily\Large,
   width=180pt,
   height=120pt,
   scale only axis=true,
   ybar,
   bar width=7.5pt,
   xtick=data,
   ylabel = {Percents},
   ymin=0,ymax=100,
   ytick={0,20,40,60,80,100},
   legend pos=north west
  ]
  \addplot[fill=blue,color=blue!60,mark=none] coordinates {(1,8.98203592814371) (2,11.377245508982) (3,14.3712574850299) (4,16.1676646706587) (5,10.1796407185629) (6,16.1676646706587) (7,22.7544910179641) };
  \addplot[fill=orange,color=orange!60,mark=none] coordinates {(1,5.45454545454545) (2,12.7272727272727) (3,14.5454545454545) (4,27.2727272727273) (5,12.7272727272727) (6,18.1818181818182) (7,9.09090909090909) };
	\addplot[sharp plot,blue,line width=1.3] coordinates {(1,8.98203592814371) (2,20.3592814371257) (3,34.7305389221557) (4,50.8982035928144) (5,61.0778443113772) (6,77.2455089820359) (7,100) };
	\addplot[sharp plot,orange,line width=1.3] coordinates {(1,5.45454545454545) (2,18.1818181818182) (3,32.7272727272727) (4,60) (5,72.7272727272727) (6,90.9090909090909) (7,100) };
  \legend{Dev.,Res.}
 \end{axis}
\end{tikzpicture}}
\\
\hline

\end{tabular}
\vspace{2mm}

%% file: plt-14.tex
\vspace{1mm}
\noindent\begin{tabular}{@{}p{45mm}p{37mm}}
\hline
\rule{0pt}{2ex}\raggedright Q14: Using the project for research with no scientific value&

\adjustbox{valign=t}{\begin{tikzpicture}[scale=0.5]
 \begin{axis}[
   font=\sffamily\Large,
   width=180pt,
   height=120pt,
   scale only axis=true,
   ybar,
   bar width=7.5pt,
   xtick=data,
   ylabel = {Percents},
   ymin=0,ymax=100,
   ytick={0,20,40,60,80,100},
   legend pos=north west
  ]
  \addplot[fill=blue,color=blue!60,mark=none] coordinates {(1,20.2453987730061) (2,9.20245398773006) (3,4.29447852760736) (4,9.8159509202454) (5,9.20245398773006) (6,13.4969325153374) (7,33.7423312883436) };
  \addplot[fill=orange,color=orange!60,mark=none] coordinates {(1,24.5283018867925) (2,5.66037735849057) (3,0) (4,7.54716981132075) (5,13.2075471698113) (6,15.0943396226415) (7,33.9622641509434) };
	\addplot[sharp plot,blue,line width=1.3] coordinates {(1,20.2453987730061) (2,29.4478527607362) (3,33.7423312883436) (4,43.558282208589) (5,52.760736196319) (6,66.2576687116564) (7,100) };
	\addplot[sharp plot,orange,line width=1.3] coordinates {(1,24.5283018867925) (2,30.188679245283) (3,30.188679245283) (4,37.7358490566038) (5,50.9433962264151) (6,66.0377358490566) (7,100) };
  \legend{Dev.,Res.}
 \end{axis}
\end{tikzpicture}}
\\
\hline
\rule{0pt}{2ex}\raggedright Q15: Using the project to test a new experimental methodology&

\adjustbox{valign=t}{\begin{tikzpicture}[scale=0.5]
 \begin{axis}[
   font=\sffamily\Large,
   width=180pt,
   height=120pt,
   scale only axis=true,
   ybar,
   bar width=7.5pt,
   xtick=data,
   ylabel = {Percents},
   ymin=0,ymax=100,
   ytick={0,20,40,60,80,100},
   legend style={at={(0.98,0.4)},anchor=east}
  ]
  \addplot[fill=blue,color=blue!60,mark=none] coordinates {(1,25.609756097561) (2,13.4146341463415) (3,10.3658536585366) (4,14.6341463414634) (5,15.2439024390244) (6,12.1951219512195) (7,8.53658536585366) };
  \addplot[fill=orange,color=orange!60,mark=none] coordinates {(1,30.9090909090909) (2,10.9090909090909) (3,9.09090909090909) (4,18.1818181818182) (5,12.7272727272727) (6,12.7272727272727) (7,5.45454545454545) };
	\addplot[sharp plot,blue,line width=1.3] coordinates {(1,25.609756097561) (2,39.0243902439024) (3,49.390243902439) (4,64.0243902439024) (5,79.2682926829268) (6,91.4634146341463) (7,100) };
	\addplot[sharp plot,orange,line width=1.3] coordinates {(1,30.9090909090909) (2,41.8181818181818) (3,50.9090909090909) (4,69.0909090909091) (5,81.8181818181818) (6,94.5454545454545) (7,100) };
  \legend{Dev.,Res.}
 \end{axis}
\end{tikzpicture}}
\\
\hline

\end{tabular}
\vspace{2mm}

%% file: plt-16.tex
\vspace{1mm}
\noindent\begin{tabular}{@{}p{45mm}p{37mm}}
\hline
\rule{0pt}{2ex}\raggedright Q16: Contributing buggy code to the project&

\adjustbox{valign=t}{\begin{tikzpicture}[scale=0.5]
 \begin{axis}[
   font=\sffamily\Large,
   width=180pt,
   height=120pt,
   scale only axis=true,
   ybar,
   bar width=7.5pt,
   xtick=data,
   ylabel = {Percents},
   ymin=0,ymax=100,
   ytick={0,20,40,60,80,100},
   legend pos=north west
  ]
  \addplot[fill=blue,color=blue!60,mark=none] coordinates {(1,1.20481927710843) (2,1.20481927710843) (3,2.40963855421687) (4,8.43373493975904) (5,6.02409638554217) (6,16.8674698795181) (7,63.855421686747) };
  \addplot[fill=orange,color=orange!60,mark=none] coordinates {(1,1.81818181818182) (2,0) (3,0) (4,3.63636363636364) (5,16.3636363636364) (6,23.6363636363636) (7,54.5454545454545) };
	\addplot[sharp plot,blue,line width=1.3] coordinates {(1,1.20481927710843) (2,2.40963855421687) (3,4.81927710843374) (4,13.2530120481928) (5,19.2771084337349) (6,36.144578313253) (7,100) };
	\addplot[sharp plot,orange,line width=1.3] coordinates {(1,1.81818181818182) (2,1.81818181818182) (3,1.81818181818182) (4,5.45454545454545) (5,21.8181818181818) (6,45.4545454545455) (7,100) };
  \legend{Dev.,Res.}
 \end{axis}
\end{tikzpicture}}
\\
\hline
\rule{0pt}{2ex}\raggedright Q17: Wasting the time of maintainers to review and reject code&

\adjustbox{valign=t}{\begin{tikzpicture}[scale=0.5]
 \begin{axis}[
   font=\sffamily\Large,
   width=180pt,
   height=120pt,
   scale only axis=true,
   ybar,
   bar width=7.5pt,
   xtick=data,
   ylabel = {Percents},
   ymin=0,ymax=100,
   ytick={0,20,40,60,80,100},
   legend pos=north west
  ]
  \addplot[fill=blue,color=blue!60,mark=none] coordinates {(1,1.80722891566265) (2,1.20481927710843) (3,6.02409638554217) (4,9.03614457831325) (5,16.8674698795181) (6,18.0722891566265) (7,46.9879518072289) };
  \addplot[fill=orange,color=orange!60,mark=none] coordinates {(1,1.81818181818182) (2,7.27272727272727) (3,0) (4,10.9090909090909) (5,14.5454545454545) (6,20) (7,45.4545454545455) };
	\addplot[sharp plot,blue,line width=1.3] coordinates {(1,1.80722891566265) (2,3.01204819277108) (3,9.03614457831325) (4,18.0722891566265) (5,34.9397590361446) (6,53.0120481927711) (7,100) };
	\addplot[sharp plot,orange,line width=1.3] coordinates {(1,1.81818181818182) (2,9.09090909090909) (3,9.09090909090909) (4,20) (5,34.5454545454545) (6,54.5454545454545) (7,100) };
  \legend{Dev.,Res.}
 \end{axis}
\end{tikzpicture}}
\\
\hline

\end{tabular}
\vspace{2mm}

%% file: plt-23.tex
\vspace{1mm}
\noindent\begin{tabular}{@{}p{45mm}p{37mm}}
\hline
\rule{0pt}{2ex}\raggedright Q23: A novice programmer contributes beginner-quality code in an attempt to build up his or her reputation&

\adjustbox{valign=t}{\begin{tikzpicture}[scale=0.5]
 \begin{axis}[
   font=\sffamily\Large,
   width=180pt,
   height=120pt,
   scale only axis=true,
   ybar,
   bar width=7.5pt,
   xtick=data,
   ylabel = {Percents},
   ymin=0,ymax=100,
   ytick={0,20,40,60,80,100},
   legend pos=north west
  ]
  \addplot[fill=blue,color=blue!60,mark=none] coordinates {(-3,0) (-2,3.04878048780488) (-1,9.14634146341463) (0,1.82926829268293) (1,32.3170731707317) (2,42.6829268292683) (3,10.9756097560976) };
  \addplot[fill=orange,color=orange!60,mark=none] coordinates {(-3,0) (-2,1.81818181818182) (-1,9.09090909090909) (0,0) (1,30.9090909090909) (2,49.0909090909091) (3,9.09090909090909) };
	\addplot[sharp plot,blue,line width=1.3] coordinates {(-3,0) (-2,3.04878048780488) (-1,12.1951219512195) (0,14.0243902439024) (1,46.3414634146341) (2,89.0243902439024) (3,100) };
	\addplot[sharp plot,orange,line width=1.3] coordinates {(-3,0) (-2,1.81818181818182) (-1,10.9090909090909) (0,10.9090909090909) (1,41.8181818181818) (2,90.9090909090909) (3,100) };
  \legend{Dev.,Res.}
 \end{axis}
\end{tikzpicture}}
\\
\hline
\rule{0pt}{2ex}\raggedright Q29: A student contributes code to learn how open source development works&

\adjustbox{valign=t}{\begin{tikzpicture}[scale=0.5]
 \begin{axis}[
   font=\sffamily\Large,
   width=180pt,
   height=120pt,
   scale only axis=true,
   ybar,
   bar width=7.5pt,
   xtick=data,
   ylabel = {Percents},
   ymin=0,ymax=100,
   ytick={0,20,40,60,80,100},
   legend pos=north west
  ]
  \addplot[fill=blue,color=blue!60,mark=none] coordinates {(-3,0) (-2,1.20481927710843) (-1,8.43373493975904) (0,5.42168674698795) (1,15.0602409638554) (2,49.3975903614458) (3,20.4819277108434) };
  \addplot[fill=orange,color=orange!60,mark=none] coordinates {(-3,0) (-2,7.27272727272727) (-1,7.27272727272727) (0,5.45454545454545) (1,20) (2,47.2727272727273) (3,12.7272727272727) };
	\addplot[sharp plot,blue,line width=1.3] coordinates {(-3,0) (-2,1.20481927710843) (-1,9.63855421686747) (0,15.0602409638554) (1,30.1204819277108) (2,79.5180722891566) (3,100) };
	\addplot[sharp plot,orange,line width=1.3] coordinates {(-3,0) (-2,7.27272727272727) (-1,14.5454545454545) (0,20) (1,40) (2,87.2727272727273) (3,100) };
  \legend{Dev.,Res.}
 \end{axis}
\end{tikzpicture}}
\\
\hline
\rule{0pt}{2ex}\raggedright Q25: A developer contributes code that caters to a specific personal need&

\adjustbox{valign=t}{\begin{tikzpicture}[scale=0.5]
 \begin{axis}[
   font=\sffamily\Large,
   width=180pt,
   height=120pt,
   scale only axis=true,
   ybar,
   bar width=7.5pt,
   xtick=data,
   ylabel = {Percents},
   ymin=0,ymax=100,
   ytick={0,20,40,60,80,100},
   legend pos=north west
  ]
  \addplot[fill=blue,color=blue!60,mark=none] coordinates {(-3,0) (-2,2.42424242424242) (-1,6.06060606060606) (0,6.66666666666667) (1,21.8181818181818) (2,55.7575757575758) (3,7.27272727272727) };
  \addplot[fill=orange,color=orange!60,mark=none] coordinates {(-3,0) (-2,5.55555555555556) (-1,12.962962962963) (0,5.55555555555556) (1,24.0740740740741) (2,44.4444444444444) (3,7.40740740740741) };
	\addplot[sharp plot,blue,line width=1.3] coordinates {(-3,0) (-2,2.42424242424242) (-1,8.48484848484848) (0,15.1515151515152) (1,36.969696969697) (2,92.7272727272727) (3,100) };
	\addplot[sharp plot,orange,line width=1.3] coordinates {(-3,0) (-2,5.55555555555556) (-1,18.5185185185185) (0,24.0740740740741) (1,48.1481481481482) (2,92.5925925925926) (3,100) };
  \legend{Dev.,Res.}
 \end{axis}
\end{tikzpicture}}
\\
\hline

\end{tabular}
\vspace{2mm}

%% file: plt-30.tex
\vspace{1mm}
\noindent\begin{tabular}{@{}p{45mm}p{37mm}}
\hline
\rule{0pt}{2ex}\raggedright Q30: A researcher analyzes open source code in research on the use of certain programming language constructs&

\adjustbox{valign=t}{\begin{tikzpicture}[scale=0.5]
 \begin{axis}[
   font=\sffamily\Large,
   width=180pt,
   height=120pt,
   scale only axis=true,
   ybar,
   bar width=7.5pt,
   xtick=data,
   ylabel = {Percents},
   ymin=0,ymax=100,
   ytick={0,20,40,60,80,100},
   legend pos=north west
  ]
  \addplot[fill=blue,color=blue!60,mark=none] coordinates {(-3,0) (-2,0.602409638554217) (-1,0) (0,2.40963855421687) (1,2.40963855421687) (2,53.0120481927711) (3,41.566265060241) };
  \addplot[fill=orange,color=orange!60,mark=none] coordinates {(-3,0) (-2,0) (-1,0) (0,0) (1,3.63636363636364) (2,61.8181818181818) (3,34.5454545454545) };
	\addplot[sharp plot,blue,line width=1.3] coordinates {(-3,0) (-2,0.602409638554217) (-1,0.602409638554217) (0,3.01204819277108) (1,5.42168674698795) (2,58.433734939759) (3,100) };
	\addplot[sharp plot,orange,line width=1.3] coordinates {(-3,0) (-2,0) (-1,0) (0,0) (1,3.63636363636364) (2,65.4545454545455) (3,100) };
  \legend{Dev.,Res.}
 \end{axis}
\end{tikzpicture}}
\\
\hline
\rule{0pt}{2ex}\raggedright Q31: A researcher contributes valid bug fixes to see how long it takes to incorporate them into the codebase&

\adjustbox{valign=t}{\begin{tikzpicture}[scale=0.5]
 \begin{axis}[
   font=\sffamily\Large,
   width=180pt,
   height=120pt,
   scale only axis=true,
   ybar,
   bar width=7.5pt,
   xtick=data,
   ylabel = {Percents},
   ymin=0,ymax=100,
   ytick={0,20,40,60,80,100},
   legend pos=north west
  ]
  \addplot[fill=blue,color=blue!60,mark=none] coordinates {(-3,1.20481927710843) (-2,0.602409638554217) (-1,1.80722891566265) (0,1.80722891566265) (1,3.6144578313253) (2,56.6265060240964) (3,34.3373493975904) };
  \addplot[fill=orange,color=orange!60,mark=none] coordinates {(-3,1.81818181818182) (-2,3.63636363636364) (-1,1.81818181818182) (0,3.63636363636364) (1,9.09090909090909) (2,52.7272727272727) (3,27.2727272727273) };
	\addplot[sharp plot,blue,line width=1.3] coordinates {(-3,1.20481927710843) (-2,1.80722891566265) (-1,3.6144578313253) (0,5.42168674698795) (1,9.03614457831325) (2,65.6626506024096) (3,100) };
	\addplot[sharp plot,orange,line width=1.3] coordinates {(-3,1.81818181818182) (-2,5.45454545454545) (-1,7.27272727272727) (0,10.9090909090909) (1,20) (2,72.7272727272727) (3,100) };
  \legend{Dev.,Res.}
 \end{axis}
\end{tikzpicture}}
\\
\hline
\rule{0pt}{2ex}\raggedright Q33: A researcher contributes code with a minor bug to see if it would be caught&

\adjustbox{valign=t}{\begin{tikzpicture}[scale=0.5]
 \begin{axis}[
   font=\sffamily\Large,
   width=180pt,
   height=120pt,
   scale only axis=true,
   ybar,
   bar width=7.5pt,
   xtick=data,
   ylabel = {Percents},
   ymin=0,ymax=100,
   ytick={0,20,40,60,80,100},
   legend style={at={(0.98,0.5)},anchor=east}
  ]
  \addplot[fill=blue,color=blue!60,mark=none] coordinates {(-3,12.6506024096386) (-2,45.7831325301205) (-1,25.9036144578313) (0,3.6144578313253) (1,9.03614457831325) (2,2.40963855421687) (3,0.602409638554217) };
  \addplot[fill=orange,color=orange!60,mark=none] coordinates {(-3,1.81818181818182) (-2,70.9090909090909) (-1,18.1818181818182) (0,1.81818181818182) (1,7.27272727272727) (2,0) (3,0) };
	\addplot[sharp plot,blue,line width=1.3] coordinates {(-3,12.6506024096386) (-2,58.433734939759) (-1,84.3373493975904) (0,87.9518072289157) (1,96.9879518072289) (2,99.3975903614458) (3,100) };
	\addplot[sharp plot,orange,line width=1.3] coordinates {(-3,1.81818181818182) (-2,72.7272727272727) (-1,90.9090909090909) (0,92.7272727272727) (1,100) (2,100) (3,100) };
  \legend{Dev.,Res.}
 \end{axis}
\end{tikzpicture}}
\\
\hline
\rule{0pt}{2ex}\raggedright Q34: A researcher contributes code with a security bug to see if it would be caught&

\adjustbox{valign=t}{\begin{tikzpicture}[scale=0.5]
 \begin{axis}[
   font=\sffamily\Large,
   width=180pt,
   height=120pt,
   scale only axis=true,
   ybar,
   bar width=7.5pt,
   xtick=data,
   ylabel = {Percents},
   ymin=0,ymax=100,
   ytick={0,20,40,60,80,100},
   legend style={at={(0.98,0.5)},anchor=east}
  ]
  \addplot[fill=blue,color=blue!60,mark=none] coordinates {(-3,43.6363636363636) (-2,36.3636363636364) (-1,10.9090909090909) (0,3.03030303030303) (1,3.63636363636364) (2,1.81818181818182) (3,0.606060606060606) };
  \addplot[fill=orange,color=orange!60,mark=none] coordinates {(-3,47.2727272727273) (-2,34.5454545454545) (-1,9.09090909090909) (0,3.63636363636364) (1,3.63636363636364) (2,0) (3,1.81818181818182) };
	\addplot[sharp plot,blue,line width=1.3] coordinates {(-3,43.6363636363636) (-2,80) (-1,90.9090909090909) (0,93.9393939393939) (1,97.5757575757576) (2,99.3939393939394) (3,100) };
	\addplot[sharp plot,orange,line width=1.3] coordinates {(-3,47.2727272727273) (-2,81.8181818181818) (-1,90.9090909090909) (0,94.5454545454545) (1,98.1818181818182) (2,98.1818181818182) (3,100) };
  \legend{Dev.,Res.}
 \end{axis}
\end{tikzpicture}}
\\
\hline

\end{tabular}
\vspace{2mm}

%% file: plt-24.tex
\vspace{1mm}
\noindent\begin{tabular}{@{}p{45mm}p{37mm}}
\hline
\rule{0pt}{2ex}\raggedright Q24: A developer contributes reasonably tested code that unknowingly still contains a bug&

\adjustbox{valign=t}{\begin{tikzpicture}[scale=0.5]
 \begin{axis}[
   font=\sffamily\Large,
   width=180pt,
   height=120pt,
   scale only axis=true,
   ybar,
   bar width=7.5pt,
   xtick=data,
   ylabel = {Percents},
   ymin=0,ymax=100,
   ytick={0,20,40,60,80,100},
   legend pos=north west
  ]
  \addplot[fill=blue,color=blue!60,mark=none] coordinates {(-3,0) (-2,0) (-1,3.6144578313253) (0,0.602409638554217) (1,12.0481927710843) (2,69.8795180722892) (3,13.855421686747) };
  \addplot[fill=orange,color=orange!60,mark=none] coordinates {(-3,0) (-2,5.45454545454545) (-1,1.81818181818182) (0,0) (1,9.09090909090909) (2,69.0909090909091) (3,14.5454545454545) };
	\addplot[sharp plot,blue,line width=1.3] coordinates {(-3,0) (-2,0) (-1,3.6144578313253) (0,4.21686746987952) (1,16.2650602409639) (2,86.144578313253) (3,100) };
	\addplot[sharp plot,orange,line width=1.3] coordinates {(-3,0) (-2,5.45454545454545) (-1,7.27272727272727) (0,7.27272727272727) (1,16.3636363636364) (2,85.4545454545455) (3,100) };
  \legend{Dev.,Res.}
 \end{axis}
\end{tikzpicture}}
\\
\hline
\rule{0pt}{2ex}\raggedright Q26: A developer contributes code that was not adequately tested&

\adjustbox{valign=t}{\begin{tikzpicture}[scale=0.5]
 \begin{axis}[
   font=\sffamily\Large,
   width=180pt,
   height=120pt,
   scale only axis=true,
   ybar,
   bar width=7.5pt,
   xtick=data,
   ylabel = {Percents},
   ymin=0,ymax=100,
   ytick={0,20,40,60,80,100},
   legend style={at={(0.98,0.6)},anchor=east}
  ]
  \addplot[fill=blue,color=blue!60,mark=none] coordinates {(-3,1.21212121212121) (-2,32.1212121212121) (-1,37.5757575757576) (0,3.03030303030303) (1,20.6060606060606) (2,4.84848484848485) (3,0.606060606060606) };
  \addplot[fill=orange,color=orange!60,mark=none] coordinates {(-3,1.81818181818182) (-2,36.3636363636364) (-1,27.2727272727273) (0,1.81818181818182) (1,29.0909090909091) (2,1.81818181818182) (3,1.81818181818182) };
	\addplot[sharp plot,blue,line width=1.3] coordinates {(-3,1.21212121212121) (-2,33.3333333333333) (-1,70.9090909090909) (0,73.9393939393939) (1,94.5454545454545) (2,99.3939393939394) (3,100) };
	\addplot[sharp plot,orange,line width=1.3] coordinates {(-3,1.81818181818182) (-2,38.1818181818182) (-1,65.4545454545455) (0,67.2727272727273) (1,96.3636363636364) (2,98.1818181818182) (3,100) };
  \legend{Dev.,Res.}
 \end{axis}
\end{tikzpicture}}
\\
\hline

\end{tabular}
\vspace{2mm}

%% file: plt-27.tex
\vspace{1mm}
\noindent\begin{tabular}{@{}p{45mm}p{37mm}}
\hline
\rule{0pt}{2ex}\raggedright Q27: A developer contributes code based on an automatic tool after verifying it manually&

\adjustbox{valign=t}{\begin{tikzpicture}[scale=0.5]
 \begin{axis}[
   font=\sffamily\Large,
   width=180pt,
   height=120pt,
   scale only axis=true,
   ybar,
   bar width=7.5pt,
   xtick=data,
   ylabel = {Percents},
   ymin=0,ymax=100,
   ytick={0,20,40,60,80,100},
   legend pos=north west
  ]
  \addplot[fill=blue,color=blue!60,mark=none] coordinates {(-3,0) (-2,3.6144578313253) (-1,6.02409638554217) (0,13.2530120481928) (1,15.0602409638554) (2,48.1927710843374) (3,13.855421686747) };
  \addplot[fill=orange,color=orange!60,mark=none] coordinates {(-3,1.81818181818182) (-2,1.81818181818182) (-1,7.27272727272727) (0,10.9090909090909) (1,7.27272727272727) (2,63.6363636363636) (3,7.27272727272727) };
	\addplot[sharp plot,blue,line width=1.3] coordinates {(-3,0) (-2,3.6144578313253) (-1,9.63855421686747) (0,22.8915662650602) (1,37.9518072289157) (2,86.144578313253) (3,100) };
	\addplot[sharp plot,orange,line width=1.3] coordinates {(-3,1.81818181818182) (-2,3.63636363636364) (-1,10.9090909090909) (0,21.8181818181818) (1,29.0909090909091) (2,92.7272727272727) (3,100) };
  \legend{Dev.,Res.}
 \end{axis}
\end{tikzpicture}}
\\
\hline
\rule{0pt}{2ex}\raggedright Q28: A developer contributes code generated by a novel tool he or she is developing to see if the tool's output is good enough already&

\adjustbox{valign=t}{\begin{tikzpicture}[scale=0.5]
 \begin{axis}[
   font=\sffamily\Large,
   width=180pt,
   height=120pt,
   scale only axis=true,
   ybar,
   bar width=7.5pt,
   xtick=data,
   ylabel = {Percents},
   ymin=0,ymax=100,
   ytick={0,20,40,60,80,100},
   legend pos=north west
  ]
  \addplot[fill=blue,color=blue!60,mark=none] coordinates {(-3,1.21212121212121) (-2,22.4242424242424) (-1,22.4242424242424) (0,17.5757575757576) (1,22.4242424242424) (2,13.3333333333333) (3,0.606060606060606) };
  \addplot[fill=orange,color=orange!60,mark=none] coordinates {(-3,0) (-2,23.6363636363636) (-1,29.0909090909091) (0,12.7272727272727) (1,14.5454545454545) (2,20) (3,0) };
	\addplot[sharp plot,blue,line width=1.3] coordinates {(-3,1.21212121212121) (-2,23.6363636363636) (-1,46.0606060606061) (0,63.6363636363636) (1,86.0606060606061) (2,99.3939393939394) (3,100) };
	\addplot[sharp plot,orange,line width=1.3] coordinates {(-3,0) (-2,23.6363636363636) (-1,52.7272727272727) (0,65.4545454545455) (1,80) (2,100) (3,100) };
  \legend{Dev.,Res.}
 \end{axis}
\end{tikzpicture}}
\\
\hline
\rule{0pt}{2ex}\raggedright Q32: A researcher contributes code suggested by an experimental tool he or she is developing to assess the tool's possible contribution&

\adjustbox{valign=t}{\begin{tikzpicture}[scale=0.5]
 \begin{axis}[
   font=\sffamily\Large,
   width=180pt,
   height=120pt,
   scale only axis=true,
   ybar,
   bar width=7.5pt,
   xtick=data,
   ylabel = {Percents},
   ymin=0,ymax=100,
   ytick={0,20,40,60,80,100},
   legend pos=north west
  ]
  \addplot[fill=blue,color=blue!60,mark=none] coordinates {(-3,0.602409638554217) (-2,9.63855421686747) (-1,15.0602409638554) (0,18.0722891566265) (1,24.6987951807229) (2,28.3132530120482) (3,3.6144578313253) };
  \addplot[fill=orange,color=orange!60,mark=none] coordinates {(-3,0) (-2,5.45454545454545) (-1,23.6363636363636) (0,14.5454545454545) (1,23.6363636363636) (2,29.0909090909091) (3,3.63636363636364) };
	\addplot[sharp plot,blue,line width=1.3] coordinates {(-3,0.602409638554217) (-2,10.2409638554217) (-1,25.3012048192771) (0,43.3734939759036) (1,68.0722891566265) (2,96.3855421686747) (3,100) };
	\addplot[sharp plot,orange,line width=1.3] coordinates {(-3,0) (-2,5.45454545454545) (-1,29.0909090909091) (0,43.6363636363636) (1,67.2727272727273) (2,96.3636363636364) (3,100) };
  \legend{Dev.,Res.}
 \end{axis}
\end{tikzpicture}}
\\
\hline

\end{tabular}
\vspace{2mm}

%% file: plt-36.tex
\vspace{1mm}
\noindent\begin{tabular}{@{}p{45mm}p{37mm}}
\hline
\rule{0pt}{2ex}\raggedright Q36: A researcher analyzes the development trajectory of the code in an open-source project, and identifies the project in the research report&

\adjustbox{valign=t}{\begin{tikzpicture}[scale=0.5]
 \begin{axis}[
   font=\sffamily\Large,
   width=180pt,
   height=120pt,
   scale only axis=true,
   ybar,
   bar width=7.5pt,
   xtick=data,
   ylabel = {Percents},
   ymin=0,ymax=100,
   ytick={0,20,40,60,80,100},
   legend pos=north west
  ]
  \addplot[fill=blue,color=blue!60,mark=none] coordinates {(-3,0) (-2,1.80722891566265) (-1,7.2289156626506) (0,11.4457831325301) (1,10.2409638554217) (2,56.0240963855422) (3,13.2530120481928) };
  \addplot[fill=orange,color=orange!60,mark=none] coordinates {(-3,1.81818181818182) (-2,0) (-1,5.45454545454545) (0,7.27272727272727) (1,12.7272727272727) (2,58.1818181818182) (3,14.5454545454545) };
	\addplot[sharp plot,blue,line width=1.3] coordinates {(-3,0) (-2,1.80722891566265) (-1,9.03614457831325) (0,20.4819277108434) (1,30.7228915662651) (2,86.7469879518072) (3,100) };
	\addplot[sharp plot,orange,line width=1.3] coordinates {(-3,1.81818181818182) (-2,1.81818181818182) (-1,7.27272727272727) (0,14.5454545454545) (1,27.2727272727273) (2,85.4545454545455) (3,100) };
  \legend{Dev.,Res.}
 \end{axis}
\end{tikzpicture}}
\\
\hline
\rule{0pt}{2ex}\raggedright Q37: A researcher analyzes the interactions among developers in a project, and identifies the project in the research report&

\adjustbox{valign=t}{\begin{tikzpicture}[scale=0.5]
 \begin{axis}[
   font=\sffamily\Large,
   width=180pt,
   height=120pt,
   scale only axis=true,
   ybar,
   bar width=7.5pt,
   xtick=data,
   ylabel = {Percents},
   ymin=0,ymax=100,
   ytick={0,20,40,60,80,100},
   legend pos=north west
  ]
  \addplot[fill=blue,color=blue!60,mark=none] coordinates {(-3,0) (-2,5.42168674698795) (-1,13.855421686747) (0,12.6506024096386) (1,12.0481927710843) (2,47.5903614457831) (3,8.43373493975904) };
  \addplot[fill=orange,color=orange!60,mark=none] coordinates {(-3,0) (-2,1.81818181818182) (-1,9.09090909090909) (0,9.09090909090909) (1,16.3636363636364) (2,52.7272727272727) (3,10.9090909090909) };
	\addplot[sharp plot,blue,line width=1.3] coordinates {(-3,0) (-2,5.42168674698795) (-1,19.2771084337349) (0,31.9277108433735) (1,43.9759036144578) (2,91.566265060241) (3,100) };
	\addplot[sharp plot,orange,line width=1.3] coordinates {(-3,0) (-2,1.81818181818182) (-1,10.9090909090909) (0,20) (1,36.3636363636364) (2,89.0909090909091) (3,100) };
  \legend{Dev.,Res.}
 \end{axis}
\end{tikzpicture}}
\\
\hline
\rule{0pt}{2ex}\raggedright Q38: A researcher analyzes commit messages, and includes examples with the identity of the developers in the research report&

\adjustbox{valign=t}{\begin{tikzpicture}[scale=0.5]
 \begin{axis}[
   font=\sffamily\Large,
   width=180pt,
   height=120pt,
   scale only axis=true,
   ybar,
   bar width=7.5pt,
   xtick=data,
   ylabel = {Percents},
   ymin=0,ymax=100,
   ytick={0,20,40,60,80,100},
   legend pos=north west
  ]
  \addplot[fill=blue,color=blue!60,mark=none] coordinates {(-3,3.01204819277108) (-2,19.2771084337349) (-1,22.289156626506) (0,9.63855421686747) (1,15.6626506024096) (2,26.5060240963855) (3,3.6144578313253) };
  \addplot[fill=orange,color=orange!60,mark=none] coordinates {(-3,5.45454545454545) (-2,40) (-1,14.5454545454545) (0,3.63636363636364) (1,18.1818181818182) (2,16.3636363636364) (3,1.81818181818182) };
	\addplot[sharp plot,blue,line width=1.3] coordinates {(-3,3.01204819277108) (-2,22.289156626506) (-1,44.5783132530121) (0,54.2168674698795) (1,69.8795180722892) (2,96.3855421686747) (3,100) };
	\addplot[sharp plot,orange,line width=1.3] coordinates {(-3,5.45454545454545) (-2,45.4545454545455) (-1,60) (0,63.6363636363636) (1,81.8181818181818) (2,98.1818181818182) (3,100) };
  \legend{Dev.,Res.}
 \end{axis}
\end{tikzpicture}}
\\
\hline

\end{tabular}
\vspace{2mm}

%% file: plt-35.tex
\vspace{1mm}
\noindent\begin{tabular}{@{}p{45mm}p{37mm}}
\hline
\rule{0pt}{2ex}\raggedright Q35: A researcher sends a questionnaire about open source development to emails of developers listed as contributors to an open source project&

\adjustbox{valign=t}{\begin{tikzpicture}[scale=0.5]
 \begin{axis}[
   font=\sffamily\Large,
   width=180pt,
   height=120pt,
   scale only axis=true,
   ybar,
   bar width=7.5pt,
   xtick=data,
   ylabel = {Percents},
   ymin=0,ymax=100,
   ytick={0,20,40,60,80,100},
   legend pos=north west
  ]
  \addplot[fill=blue,color=blue!60,mark=none] coordinates {(-3,2.40963855421687) (-2,3.01204819277108) (-1,13.855421686747) (0,4.21686746987952) (1,22.289156626506) (2,46.9879518072289) (3,7.2289156626506) };
  \addplot[fill=orange,color=orange!60,mark=none] coordinates {(-3,0) (-2,1.81818181818182) (-1,14.5454545454545) (0,10.9090909090909) (1,27.2727272727273) (2,36.3636363636364) (3,9.09090909090909) };
	\addplot[sharp plot,blue,line width=1.3] coordinates {(-3,2.40963855421687) (-2,5.42168674698795) (-1,19.2771084337349) (0,23.4939759036145) (1,45.7831325301205) (2,92.7710843373494) (3,100) };
	\addplot[sharp plot,orange,line width=1.3] coordinates {(-3,0) (-2,1.81818181818182) (-1,16.3636363636364) (0,27.2727272727273) (1,54.5454545454545) (2,90.9090909090909) (3,100) };
  \legend{Dev.,Res.}
 \end{axis}
\end{tikzpicture}}
\\
\hline

\end{tabular}
\vspace{2mm}